\documentclass[reprint,twocolumn]{revtex4-2}
\usepackage{graphicx}
\usepackage{amsmath,amssymb,graphicx,bm}

\newcommand{\OV}{\overline V}

\begin{document}

\title{Hamiltonian perspective on parquet theory}

\author{Frederick Green$^1$ and Thomas L. Ainsworth$^2$}
\affiliation{
$^1$ School of Physics, The University of New South Wales,
Sydney, NSW 2052, Australia\\
$^2$ Department of Physics, Kent State University, Kent, OH 44240, USA  
}
%



%
\begin{abstract}
Understanding collective phenomena calls for tractable descriptions of
correlations in assemblies of strongly interacting constituents. Capturing
the essence of their self-consistency is central. The parquet theory admits
a maximum level of self-consistency for strictly pairwise many-body
correlations. While perturbatively based, the core of parquet and allied
models is a set of strongly coupled nonlinear integral equations for
all-order scattering; tightly constrained by crossing symmetry, they are
nevertheless heuristic. Within a formalism due to Kraichnan, we present a
Hamiltonian analysis of fermionic parquet's structure. The shape of its
constitutive equations follows naturally from the resulting canonical
description. We discuss the affinity between the derived conserving
scattering amplitude and that of standard parquet. Whereas the
Hamiltonian-derived model amplitude is microscopically conserving, it
cannot preserve crossing symmetry. The parquet amplitude and its
refinements preserve crossing symmetry, yet cannot safeguard conservation
at any stage. Which amplitude should be used depends on physics rather
than on theoretically ideal completeness.
\end{abstract}


%
\maketitle

\section{Introduction}

In this paper we explore a canonical basis for a
significant class of theories of many-particle correlations.
Growth in computing capacity has fueled increasingly comprehensive studies
of assemblies of interacting elements and how these come to determine
the behavioral complexity of such assemblies, at simulational and
analytical levels. The resulting numerics feed back to expand theoretical
concepts of how a system's elementary components, with their interactions,
co-operate in subtle collective phenomena.

Among long-established formulations are
the conserving $\Phi$-derivable approximations after Kadanoff and Baym
\cite{kb1,kb2}
and an especially significant candidate, parquet theory.
\cite{pqt1,pqt2,pqt2a,pqt3}
Our program also covers parquet-like variants such as the
induced interaction
\cite{bb1,bb2,bb3}
which has a successful record in its own right for problems of strong
correlations. There is conceptual merit in codifying
the intuition behind these heuristic models in a more top-down way.

$\Phi$ derivability concerns the response structure that emerges from
constructing, as its generator, an effective correlation energy
functional $\Phi$.
In parquet one constructs the correlated two-body
scattering amplitude directly.
The interrelationship of parquet and $\Phi$ derivability
has been analyzed previously
\cite{js,roger,janis1,janis2}
though not from a Hamiltonian point of view.

Parquet, its relatives and the $\Phi$-derivable descriptions
are all constructed
by choosing judiciously, if by hand, physically dominant substructures
out of the complete set of correlation energy diagrams.
Parquet theory stands out by including topologically the largest
conceivable set of particle-particle-only and particle-hole-only
pair scattering processes. This maximally pair-coupled topology
makes it worth seeking canonical grounds for parquet to shed a different
light on its structure.

Emerging from a formalism relatively unfamiliar to many-body practice,
our conclusions turn out to resonate strongly with the
diagrammatic investigation by Smith.
\cite{roger}
To establish a Hamiltonian basis for the class of theories in question,
we adapt the strategy originally devised by Kraichnan
\cite{k1,k2}
and applied recently to a series of simpler
self-consistent diagrammatic models.
\cite{KI}
These are $\Phi$-derivable in the sense of Baym-Kadanoff;
they each possess a model Luttinger-Ward-like
correlation energy functional
\cite{lw}
as the generator of static and dynamic response and correlation functions
which, while approximate, strictly conserve particle number,
momentum and energy at both microscopic and global levels.

In essential form the parquet scattering amplitude is subsumed
under a specific $\Phi$-derivable description, the
fluctuation-exchange (FLEX) approximation.
\cite{pqt3}
Although considered incomplete and subject to refinement, the simplest
configuration of parquet is thus already a component of a correlated
model whose desirable microscopic conservation properties follow naturally.
What is not in place is a Hamiltonian underpinning for FLEX.

To arrive at any conserving response formulation a price is paid in
committing to a canonical Hamiltonian description. Granted all its
consequent analytical benefits, there is a caveat on the possibility of
further consistent refinement of parquet beyond its basic form emerging
directly from FLEX: diagrammatic iteration of the renormalized two-body
parquet amplitude, feeding it back into the one-body self-energy
functional, cannot achieve control over conservation.
\cite{roger}
We revisit this in the following.

Kraichnan's embedding of the many-body problem in a larger space departs
substantially from traditional diagrammatic reasoning. It is central to the
project because it is applicable to systems with pair interactions. In
principle it should have something to say about parquet.

One starts by injecting the physical Hamiltonian into a much larger sum
(going to infinity) of identical but distinguishable replicas. A collective
representation is introduced over this total Hamiltonian. Next,
isomorphic copies of these large collective Hamiltonians are themselves
summed into a grand Hamiltonian, but with the interaction potential of
each collective copy now partnered by an individual coupling factor.
The factor depends only on the collective indices as $V$ depends
only on the physical indices. Adjoined to $V$ in this way,
the coupling can be defined stochastically.

Provided the couplings transform in their abstract indices as the
elementary pair potential transforms in its physical indices, the end
result is a Hamiltonian in which the original physical form is embedded.
Left unmodified, with all coupling factors set to unity, the expanded
system recovers the exact physics. If modified appropriately, all the
unitary properties of the collective Hamiltonians, and of their grand sum,
are unaffected.

Contingent upon the functional form of the couplings, any operator
expectation over their distribution in the super-assembly allows subsets
of the exact correlation-energy diagrams at any order to survive
when their overall coupling-factor product works out as unity,
thus imparting immunity to taking the expectation. All other products of
random coupling factors are asymptotically killed off by mutually
destructive interference in the expectation: an extension
of the random-phase approximation.
\cite{dpelem}

The key to the strategy is that, up to averaging over Kraichnan's couplings,
the super-ensemble represents a well defined many-body Hamiltonian. Each
collective member is distinguished by its own assignment of coupling factors
and corresponds to a precisely defined Fock space. This means that any exact
identity in the hierarchy of analytic Green functions will survive averaging
if (and only if) the averaging process is done consistently on both sides
of the relation. This covers the Ward-Pitaevsky identities between one-body
self-energy and two-body response kernel, and Kramers-Kr\"onig analyticity
leading to the frequency sum rules for the correlation functions.
\cite{pn}
Relations that do not rely on analyticity are not preserved, however.
We clarify the distinction in the following.

One is therefore justified in discussing a canonical Hamiltonian for the
diagrammatic approximation giving the expectation for $\Phi$ over the
distribution of Kraichnan's coupling factors.
To cite Baym:
\cite{kb2}
``One reason underlying the fact that these approximations have such a
remarkable structure has been discovered by Kraichnan, who has shown that
in a certain sense they are exact solutions to model Hamiltonians
containing an infinite number of stochastic parameters''.

Unlike in a physically guided, constructive
$\Phi$-derivable model, the approximation is encoded here
{\em a priori} in the couplings of the Kraichnan
Hamiltonian. This does not mean ``from first principles'', as
the intuitive task of isolating dominant terms is merely shifted
from the choice of a diagram subset for $\Phi$ to that of an
appropriate Kraichnan coupling (K-coupling hereafter).
It really means that classes of conserving consistency properties,
though not all, that are fundamental in the canonical
description hold automatically after averaging.

Section II starts with a minimal review of $\Phi$ derivability,
recalling properties essential in building up a conserving
many-body expansion. Kraichnan's formalism
\cite{k1,k2,KI}
is then introduced and a form of it proposed,
including all possible pairwise-only interactions.
In Sec. III we revisit the logical development of the pairwise
correlation structure of the Kraichnan model's response
to a perturbation. This teases out real physical effects otherwise
dormant, or virtual, in the self-consistent structure of $\Phi$ itself.
Finally in Sec. IV we arrive at the parquet equations' scaffold,
displaying its provenance from the Hamiltonian defined after Kraichnan.
There too we discuss conceptual points of difference
between parquet topology interpreted within the Hamiltonian outlook,
and attempts to enlarge the topology by an iterative feedback;
these do not accord with $\Phi$ derivability.

Despite their affinity, $\Phi$ derivability and parquet analysis
exhibit complementary inherent shortcomings deeply linked to the
general nature of so-called planar diagrammatic expansions.
\cite{roger,js}
This invites care in considering which sets of physical problems
are better served by one or other of the two approaches.
We offer concluding observations in Sec. V.

\section{Precise Hamiltonians for Approximate Models}

\subsection{Correlation energy}

Our many-body system has the second-quantized Hamiltonian
\begin{eqnarray}
{\cal H}
&=&
\sum_k \varepsilon_k a^*_k a_k
\cr
&&
+ {1\over 2}
{\sum_{k_1 k_2 k_3 k_4}}\!\!\!\!\!'
{\langle k_1 k_2 | V | k_3 k_4 \rangle}
a^*_{k_1} a^*_{k_2} a_{k_3} a_{k_4};
\cr
\cr
&&
{\langle k_1 k_2 | V | k_3 k_4 \rangle}
\equiv
\delta_{s_1 s_4} \delta_{s_2 s_3} 
V({\bf k}_1 - {\bf k}_4)
\label{kII01}
\end{eqnarray}
in terms of one-particle creation operators $a^*$ and
annihilation operators $a$. The first right-hand term is the usual
total kinetic energy, the second term is the pairwise interaction.
For simplicity we discuss a spin- (or isospin-) independent
scalar $V$ but this can be relaxed
without invalidating the argument for pair interactions.
Here, again for simplicity, we address a spatially uniform system
for which momentum is a good quantum number; index $k$ stands for the
wave vector and spin pair $({\bf k},s)$,
writing $a^*_k$ as the creation operator
with $a_k$ the annihilation operator, both satisfying fermion
anticommutation. The summation ${\sum}'_{k_1 k_2 k_3 k_4}$
comes with the restriction $k_1+ k_2 = k_3 + k_4$. In a
neutral uniform Coulomb system,
potential terms with $k_2 - k_3 = 0 = k_4 - k_1$
are canceled by the background and are excluded.

The ground-state energy resulting from the full Hamiltonian includes
a correlation component $\Phi[V]$, the essential generator for the
diagrammatic expansions that act as vocabulary to the grammar
of the analysis. Here we go directly to $\Phi[V]$ and for full discussion
of the interacting ground-state structure we refer to
the classic literature.
\cite{kb2,lw}
The correlation energy can always be
written as a coupling-constant integral
\begin{eqnarray}
\Phi[V] \equiv {1\over 2}\int^1_0 {dZ\over Z}
G[Z V]\!:\!\Lambda[Z V; G]\!:\!G[Z V]
\label{kII02}
\end{eqnarray}
in which $G[V]$ is the complete renormalized two-point Green function of the
system, describing propagation of a single particle in the presence of
all the rest, and $\Lambda[V; G]$ is the fully renormalized four-point
scattering amplitude whose internal structure manifests all the possible
modes by which the propagating particles (via $G$) interact via $V$.
Single dots ``$\cdot$'' and double dots ``:'' denote single and double
internal integrations respectively, over frequency, spin and wave vector,
rendering $G\!:\!\Lambda\!:\!G$ an energy expectation value.

The renormalized Green function satisfies Dyson's equation
\begin{eqnarray}
G[V]
&=&
G^{(0)} + G^{(0)}\!\cdot\! \Sigma[V;G]\!\cdot\! G[V]
\label{kII03}
\end{eqnarray}
with $G^{(0)}$ as the noninteracting Green function
$G^{(0)}_k(\omega) \equiv (\omega - \varepsilon_k)^{-1}$
with $\Sigma[V;G]$ as the self-energy.
Equation (\ref{kII03}) links back to the correlation-energy functional
self-consistently through the variation that defines the self-energy
\begin{eqnarray}
\Sigma[V;G]
&\equiv&
\frac{\delta \Phi[V]}{\delta G[V]} = \Lambda[V; G]\!:\! G[V].
\label{kII04}
\end{eqnarray}

We recall the basic requirements on $\Lambda$. In the
expansion to order $n$ in $V$ within any particular linked structure
of $G\!:\!\!\Lambda\!\!:\!G$ reduced to its bare
elements, there will be $2n$ bare propagators.
The integral effectively treats each $G^{(0)}$ as distinguishable, and
there is a $2n$-fold ambiguity as to which bare propagator should be
the seed on which the given contribution is built up. That is, integration
replicates the same graph $2n$ times from any particular $G^{(0)}$
in the integral; but the structure contributes once only in $\Phi$. The
coupling-constant formula removes the multiplicity to all orders.

The essential feature of $\Lambda$ in the exact correlation energy
functional is the following symmetry: consider the skeleton
$G^{(0)}\!:\!\!\Lambda[V; G^{(0)}]\!\!:\!G^{(0)}$.
Removal of any $G^{(0)}$ from the skeleton,
{\em at any order in} $V$, must result in the same unique
variational structure; all lines are equivalent, The same applies
when all bare lines are replaced with dressed ones.
\cite{kb1}
This is due to unitarity and ultimately to the Hermitian character
of the Hamiltonian. It also follows that $\Lambda$ must be pairwise
irreducible: removing any two propagators $G$ from $G\!:\!\Lambda\!:\!G$
cannot produce two unlinked self-energy insertions, or else there
would be inequivalent $G$s in a contribution of form
$G\!:\!\!\Lambda_1\!\!:\!GG\!:\!\!\Lambda_2\!\!:\!G$.
These conditions impose a strongly restrictive graphical structure
upon the four-point scattering kernel entering into the self-energy.

\subsection{$\Phi$ derivability}

Other than the generic symmetry of $G$ in $\Phi$, the
variational relationships among $\Lambda$, $\Sigma$ and $G$ do not
depend on topological specifics. Those relationships 
were thus adopted as defining criteria by Baym and Kadanoff
\cite{kb1,kb2}
for constructing conserving approximations: the $\Phi$-derivable models.
Choosing a subset of skeleton diagrams from the full
$\Phi[V]$ with every $G^{(0)}$ topologically equivalent and
replacing these with dressed lines guarantees unitarity
of the effective model $\Lambda$ and secures microscopic conservation
not only at the one-body level but also for the pairwise dynamic
particle-hole response under an external perturbation.

$\Phi$ derivability necessarily entails an infinite-order approximation to
the correlation structure in terms of the bare potential. While a finite
choice of skeleton diagrams of $\Phi$ fulfills formal conservation, it must
still lead to an infinite nesting of bare interactions linked by pairs of
renormalized $G$s. Self-consistency in Eqs. (\ref{kII03}) and (\ref{kII04})
is a fundamental feature of all $\Phi$-derivable models.

\subsection{Kraichnan Hamiltonian}

The authoritative references for Kraichnan Hamiltonians
are the original papers of Kraichnan.
\cite{k1,k2}
Here we follow the more recent paper by one of us, hereafter called KI.
\cite{KI}
As per the Introduction, Kraichnan's construction proceeds by
two ensemble-building steps. First, one generates an assembly of $N$
functionally identical distinguishable copies of the exact Hamiltonian,
Eq. (\ref{kII01}).
The total Hamiltonian is
\begin{eqnarray}
{\cal H}_N
&=&
\sum^N_{n=1} \sum_k \varepsilon_k a^{*(n)}_k a^{(n)}_k
\cr
&&
+ {1\over 2} \sum^N_{n=1} {\sum_{k_1 k_2 k_3 k_4}}\!\!\!\!\!'
{\langle k_1 k_2 | V | k_3 k_4 \rangle}
a^{*(n)}_{k_1} a^{*(n)}_{k_2} a^{(n)}_{k_3} a^{(n)}_{k_4};
~~~ ~~
\label{kII05}
\end{eqnarray}
the creation and annihilation operators 
with equal index $n$ anticommute as normal;
for values of $n$ that differ, they commute.
At this point one goes over to a collective description of
the $N$-fold ensemble by Fourier transforming over index $n$.
For integer $\nu$ define the collective operators
\begin{widetext}
\begin{eqnarray}
a^{*[\nu]}_k
&\equiv&
N^{-1/2} \sum^N_{n=1} e^{2\pi i \nu n/N} a^{*(n)}_k
~~ {\rm and}~~ 
a^{[\nu]}_k
\equiv
N^{-1/2} \sum^N_{n=1} e^{-2\pi i \nu n/N} a^{(n)}_k.
\label{kII06}
\end{eqnarray}
These preserve anticommutation up to a term
strongly suppressed by mutual interference among unequal phase factors
and at most of vanishing order $1/N$.
The argument is a random-phase one:
\cite{dpelem}
\begin{eqnarray}
[a^{*[\nu]}_k, a^{[\nu']}_{k'}]_+
&=&
\frac{1}{N} \sum_{n,n'} e^{2\pi i(\nu n - \nu' n')/N}
[a^{*(n)}_k, a^{(n')}_{k'}]_+
\cr
&=&
\frac{1}{N} \sum_n e^{2\pi i(\nu - \nu')n/N} [a^{*(n)}_k, a^{(n)}_{k'}]_+
+ \frac{2}{N} \sum_{n \neq n'} e^{2\pi i(\nu n - \nu' n')/N}
a^{*(n)}_k a^{(n')}_{k'}
\cr
&=&
\delta_{kk'}\delta_{\nu\nu'} + {\cal O}(N^{-1}).
\label{kII07}
\end{eqnarray}
Similarly for $[a^{[\nu]}_k, a^{[\nu']}_{k'}]_+$. In practice, any
term in the Wick expansion of physical expectations that links
dissimilar elements $n \neq n'$ will not contribute in any case.
%
The transformation yields the new representation
\begin{eqnarray}
{\cal H}_N
&=&
\sum^N_{\nu=1} \sum_k \varepsilon_k a^{*[\nu]}_k a^{[\nu]}_k
+ {1\over 2N} {\sum_{k_1 k_2 k_3 k_4}}\!\!\!\!\!' ~~
\sum^N_{\nu_1 \nu_2 \nu_3 \nu_4}
\delta_{\nu_1+\nu_2, \nu_3+\nu_4}
~{\langle k_1 k_2 | V | k_3 k_4 \rangle}~
a^{*[\nu_1]}_{k_1} a^{*[\nu_2]}_{k_2} a^{[\nu_3]}_{k_3} a^{[\nu_4]}_{k_4}
\cr
&\equiv&
\sum_{\ell} \varepsilon_k a^*_{\ell} a_{\ell}
+ 
{1\over 2N} {\sum_{\ell_1 \ell_2 \ell_3 \ell_4}}\!\!\!\!'
~{\langle k_1 k_2 | V | k_3 k_4 \rangle}~
a^*_{\ell_1} a^*_{\ell_2} a_{\ell_3} a_{\ell_4}
\label{kII08}
\end{eqnarray}
where in the last right-hand expression we condense the notation so
$\ell \equiv (k, \nu)$ and the restriction on the sum now comprises
$\nu_1 + \nu_2 = \nu_3 + \nu_4$ (modulo $N$) as well as the
constraint on the momenta; equivalently, $\ell_1 + \ell_2 = \ell_3 + \ell_4$.

\subsection{Modifying the Hamiltonian}

\centerline{
\includegraphics[height=4truecm]{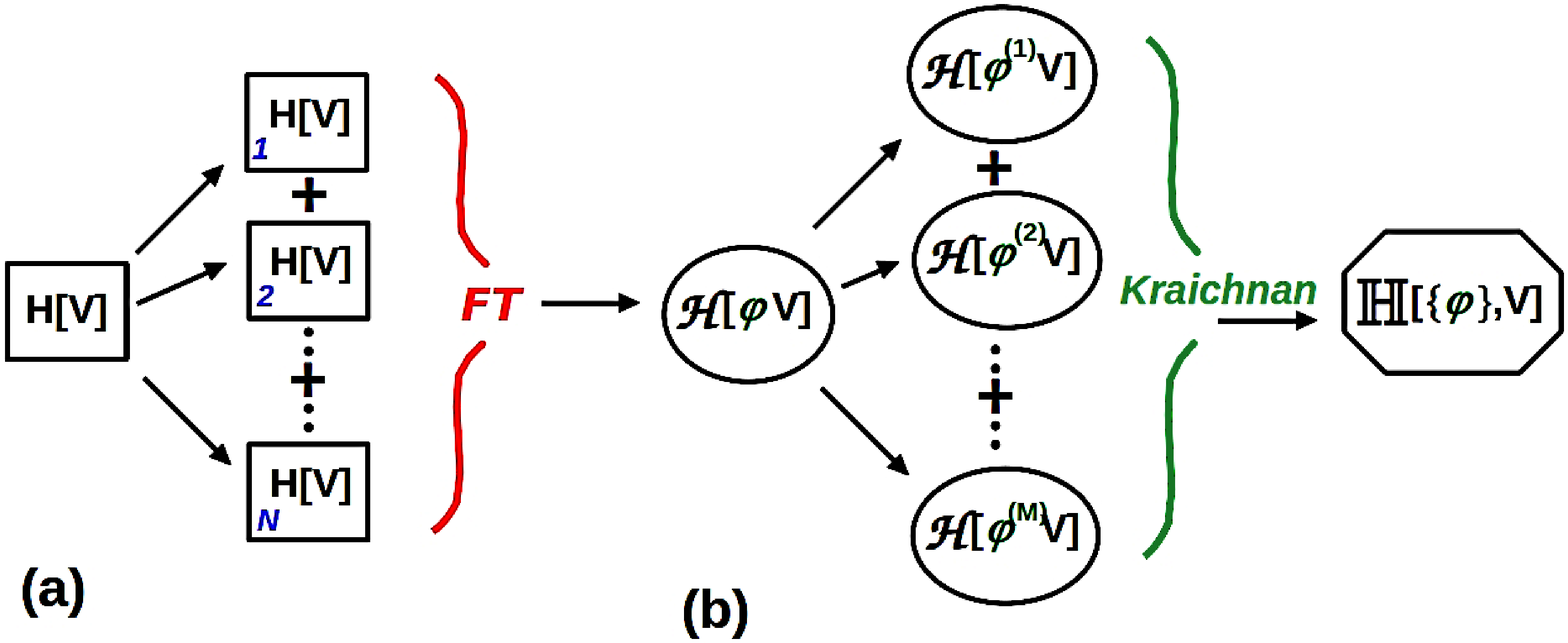}
}
{{\bf FIG. 1.} {\small Construction of the Kraichnan Hamiltonian.
(a) The exact many-body Hamiltonian is embedded in a large sum of $N$
identical but distinguishable duplicates. A Fourier transform over the
identifying index $n = 1, 2, ... N$ is performed. To each physical
interaction potential ${\langle k_1k_2|V|k_3k_4 \rangle}$ a new parameter
$\varphi_{\nu_1\nu_2|\nu_3\nu_4}$ is attached, labeled by the new Fourier
indices and transforming in them as does $V$ in its physical indices.
(b) The modified Hamiltonian is again embedded in a large sum of $M$
replicas, but each replica is now assigned a unique set of
factors $\varphi$. The resulting Kraichnan Hamiltonian remains Hermitian.
Setting every instance of $\varphi$ to unity recovers the exact
expectations resulting from the original, physical Hamiltonian. If values are
specifically structured but otherwise randomly assigned to the $M$-fold
ensemble $\{\varphi\}$, only a selected subset of the physical
correlations survives while the rest are suppressed by random phasing.
}}
\vskip 0.25cm
\end{widetext}

Performing averages given the extended Kraichnan Hamiltonian
of Eq. (\ref{kII08}), as it stands, simply recovers the exact
expectations for the originating one; any cross-correlations between
distinguishable members are identically zero. However, embedding the
physical Hamiltonian within a collective description opens a novel
degree of freedom for treating interactions. Figure 1 summarizes the
whole process. From now on we concentrate on the interaction part of
Eq. (\ref{kII08}), denoted by ${\cal H}_{i;N}$,
since the one-body part conveys no new information.
We will not consider issues of convergence here; for particular
implementations they are carefully discussed in the original papers.
\cite{k1,k2}

We can modify the
behavior of the interaction part, keeping it Hermitian, by
adjoining a factor $\varphi_{\nu_1\nu_2|\nu_3\nu_4}$ such that
\begin{eqnarray}
{\cal H}_{i;N}[\varphi]
&\equiv&
{1\over 2N} {\sum_{\ell_1 \ell_2 \ell_3 \ell_4}}\!\!\!\!'
~{\langle k_1 k_2 | V | k_3 k_4 \rangle}
~\varphi_{\nu_1\nu_2|\nu_3\nu_4}\cr
&&
~~~ ~~~ ~~~ ~~~ ~~~ ~~~ 
\times
a^*_{\ell_1} a^*_{\ell_2} a_{\ell_3} a_{\ell_4}.
\label{kII09}
\end{eqnarray}
The expression remains Hermitian if and only if the factor has the same
symmetry properties  as the potential under exchange of its indices. Thus
\begin{eqnarray}
\varphi_{\nu_4\nu_3|\nu_2\nu_1}
=
\varphi^*_{\nu_1\nu_2|\nu_3\nu_4};
~~~
\varphi_{\nu_2\nu_1|\nu_4\nu_3}
=
\varphi_{\nu_1\nu_2|\nu_3\nu_4}.
\label{kII10}
\end{eqnarray}
The additional procedure of taking expectations
will no longer match those for the exact physical Hamiltonian
unless, evidently, $\varphi$ is unity.
The crux, however, is that all identities among
expectations, dependent on causal analyticity,
will still be strictly respected.
The Kraichnan Hamiltonian remains well formed in its own right
(its Fock space is complete)
except that now it describes an abstract system necessarily
different from the physical one that motivated it. The task is
to tailor it to recover the most relevant aspects of the real
physics in reduced but tractable form.

The last step in the logic considers the much larger sum $\mathbb{H}$
of collective Hamiltonians all of the form of Eq. (\ref{kII09}),
with an interaction part $\mathbb{H}_i$ encompassing a
distribution $\{\varphi\}$ of coupling factors prescribed by a common rule:
\begin{eqnarray}
\mathbb{H}_i
\equiv
\sum_{\{\varphi\}} {\cal H}_{i;N}[\varphi].
\label{grh}
\end{eqnarray}
In Eq. (\ref{grh}) the sum ranges over the prescribed couplings.
Each Hamiltonian in the family is Hermitian,
so $\mathbb{H}$ must be also. As stated, all physical quantities -- with
one exception -- preserve their canonical
interrelationships as their expectations run through $\varphi$.

The exception is for identities relying explicitly on the completeness
of Fock space associated with $\mathbb{H}$; for, Kraichnan's ensemble
averaging destroys completeness owing to its decohering action.
Consider, symbolically, the ensemble projection operator
\begin{eqnarray*}
\mathbb{P}
\equiv \prod_{\{\varphi\}} \sum_{\Psi[\varphi]} \Psi[\varphi] \Psi^*[\varphi].
\end{eqnarray*}
Overwhelmingly the orthonormal eigenstates of $\mathbb{H}$
will be products of correlated, highly entangled, Kraichnan-coupled
superpositions of states in the Fock space of each collective member
over the distribution $\{\varphi\}$.
Any expectation over the K-couplings, directly for $\mathbb{P}$, will
cause only those terms to survive whose components in every factor
$\varphi_{\nu_1\nu_2|\nu_3\nu_4}$ within $\Psi[\varphi]$
find a counterpart in $\Psi^*[\varphi]$;
see Eq. (\ref{kII12}) below for the structure of $\varphi$.
Any other legitimate but off-diagonal cross-correlations interfere mutually
and are suppressed.
Numerically, the integrity of the Kraichnan projection operator
$\mathbb{P}$ is not preserved.
\cite{nonan}

Among other things, this loss of coherence leads to
a clear computational distinction, within the same
$\Phi$-derivable approximation,
of static (instantaneous) correlation functions over against dynamic ones.
Given that distinction, the dynamic and static response functions will
still keep their canonical definitions and the sum-rule relations among
them are still preserved.
\cite{KI,fgetal}

To align the forthcoming presentation to the notion of crossing symmetry
\cite{zqt}
for fermion interactions, we take the further
step of antisymmetrizing the potential $V$. This is readily done
in the interaction Hamiltonian, which now reads
\begin{eqnarray}
{\cal H}_{i;N}[\varphi]
&\equiv&
\!\!{1\over 2N}\! {\sum_{\ell_1 \ell_2 \ell_3 \ell_4}}\!\!\!\!'
\!\varphi_{\nu_1\nu_2|\nu_3\nu_4}
{\langle k_1 k_2 | \OV | k_3 k_4 \rangle}
a^*_{\ell_1} a^*_{\ell_2} a_{\ell_3} a_{\ell_4}
\cr
&&
\label{kII11}
\end{eqnarray}
where\[
{\langle k_1 k_2 | \OV | k_3 k_4 \rangle}
\equiv
\frac{1}{2}
( {\langle k_1 k_2 | V | k_3 k_4 \rangle}
- {\langle k_2 k_1 | V | k_3 k_4 \rangle} ).
\]
Invocations of the pair potential will now refer to Eq. (\ref{kII11}).
Care has to be taken with signs for composite ``direct'' and
``exchange'' objects that turn out actually to be mixtures of both
(yet still needing to be topologically distinguished), to make sure
the accounting for $V$ itself stays consistent.

\centerline{
\includegraphics[height=4truecm]{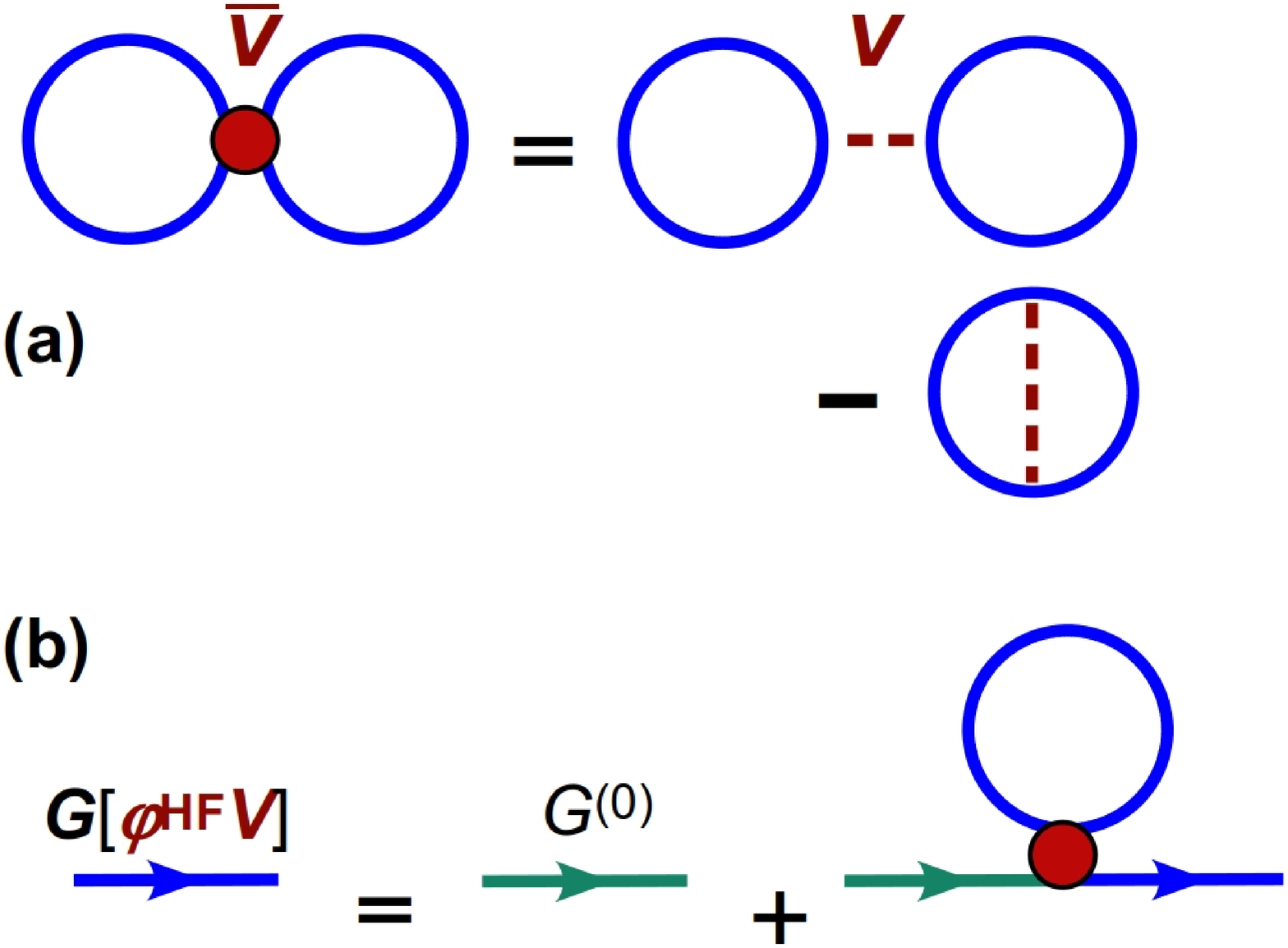}
}
{{\bf FIG. 2.} {\small Scheme for the self-consistent Hartree-Fock
interaction energy, derived by insertion into Eq. (\ref{kII11}) of
the Kraichnan coupling $\varphi^{\rm HF}_{\nu_1\nu_2|\nu_3\nu_4}
\equiv \delta_{\nu_1 \nu_4} \delta_{\nu_2 \nu_3}$. Dots: the
antisymmetrized pair interaction. Broken lines: the originating
potential. Full lines: one-body propagators.
(a) Contributions to the interaction energy. (b) Self-consistency
is made evident in the Dyson equation for the single-particle
propagator $G$, where $G^{(0)}$ is the noninteracting counterpart.
Nesting of $G\!:\!\OV$ in the self-energy contribution means that
the bare potential is present to all orders, albeit as a highly reduced
subset of the physically exact self-energy.
}}
\vskip 0.25cm

We end the review of the Kraichnan formalism by
recalling the simplest example for $\varphi$ generating the
exchange-corrected random-phase, or Hartree-Fock, approximation.
This choice is
$\varphi^{\rm HF}_{\nu_1\nu_2|\nu_3\nu_4}
\equiv \delta_{\nu_1 \nu_4} \delta_{\nu_2 \nu_3}$.
\cite{KI}
The diagrammatic outcome
of this non-stochastic Ansatz is illustrated in Figure 2. The expectation
${\langle {\cal H}_{i;N}[\varphi^{\rm HF} V] \rangle}$
of the interaction energy over $\varphi^{\rm HF}$ consists, almost trivially,
of a pair of one-body Hartree-Fock Green functions attached
to a single node representing $\OV$.

The physically richer stochastic definitions of $\varphi$,
due originally to Kraichnan
\cite{k1,k2}
were generalized and adapted in KI.
\cite{KI}
They will again be used in the next Section to build up a Kraichnan
Hamiltonian for the parquet-generating correlation energy
functional and all objects derived from it variationally.

\section{A Hamiltonian for Parquet}

\subsection{The channels and their couplings}

In simplest form, the parquet equations take the bare interaction
$\OV$ and from it build up all possible iterations that
require propagation of pairs of particles from one interaction to the next.
This excludes any contributions to the interaction energy functional
in which no interaction nodes are directly linked by such a pair;
they cannot be broken down into simpler particle-pair processes.
Two examples are shown in Fig. 3. We comment later on how these can always
be added legitimately but {\em ad hoc} to the minimal $\Phi$ functional
of immediate interest.
\centerline{
\includegraphics[height=3truecm]{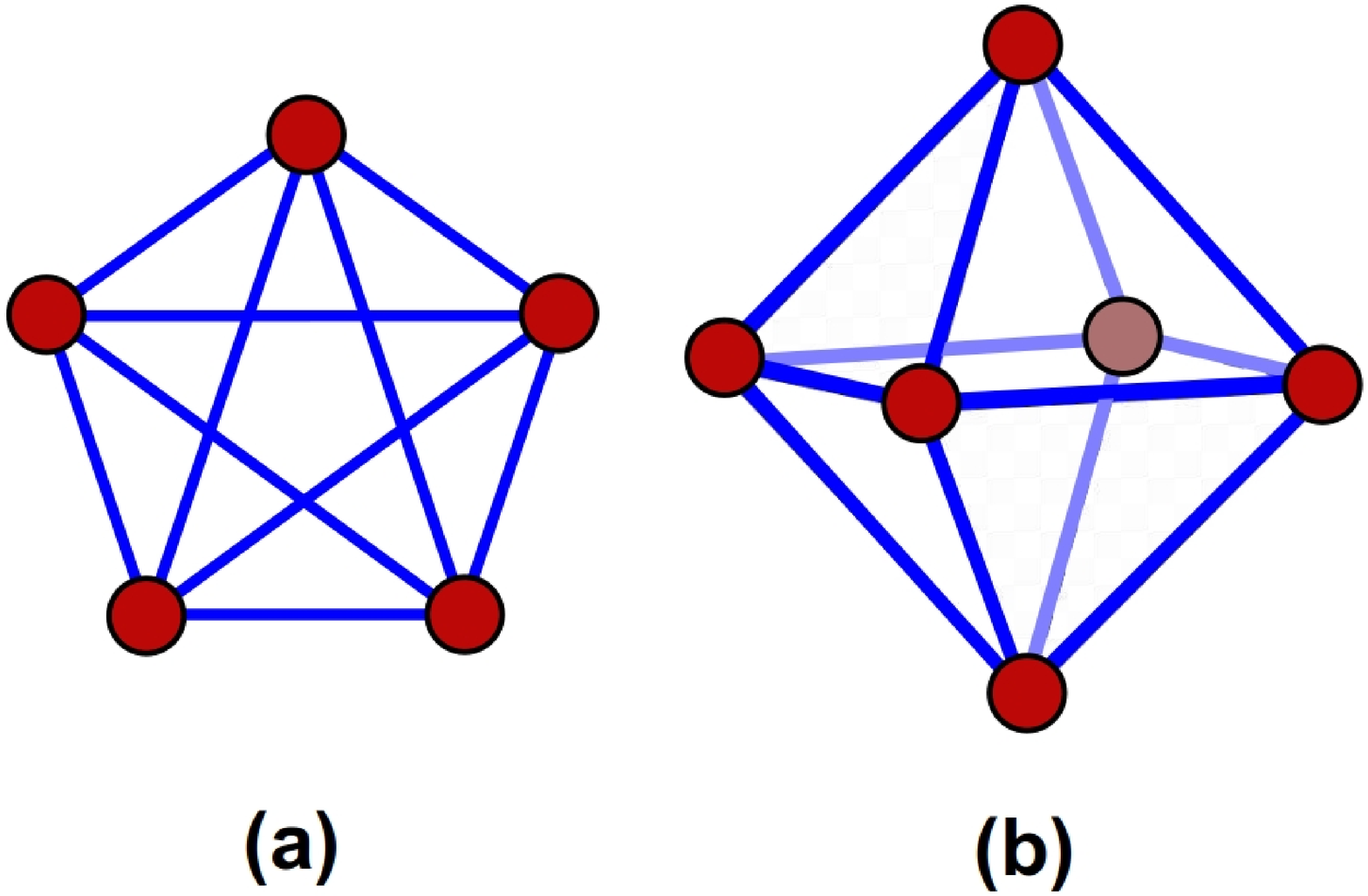}
}
{{\bf FIG. 3.} {\small Two skeleton diagrams for the exact correlation
energy not reducible to pairwise-only propagation.
(a) Next-order component, beyond first and second in $\OV$,
fulfilling the symmetry for $\Phi$ derivability
but with no two nodes directly linked by a pair of single-particle
propagators. (b) Next-higher-order term.
Such terms are not generated by any Kraichnan formulation of the
pairwise-only parquet Hamiltonian but can be added freely,
albeit only {\em ad hoc}, to its $\Phi$-derivable functional.
}}
\vskip 0.25cm

There are three possible choices of randomized K-couplings for $\varphi$,
each corresponding to the three channels included in parquet:
the $s$ channel explicitly selects propagation of pairs of particles,
the $t$ channel covers particle-hole pair propagation associated with
long-range screening in the random-phase approximation,
and its complement the $u$ channel describes the Hartree-Fock-like
exchange counterpart to $t$. Recall that while
antisymmetrization of the bare potential from $V$ to $\OV$
superposes the actual $t$ and $u$ contributions, one has to continue
distinguishing their diagrammatic representations
topologically (and their relative sign) to
preserve the quantitative outcomes of the Hamiltonian, Eq. (\ref{kII11}).

For each possible channel we define a stochastic coupling:
\begin{eqnarray}
s {\rm ~channel:~~}
\sigma _{\nu_1 \nu_2 | \nu_3 \nu_4}
&\equiv&
\exp[\pi i(\xi_{\nu_1 \nu_2} - \xi_{\nu_3 \nu_4})];
\cr
~~~ ~~~ ~~~ 
\xi_{\nu \nu'} \in [-1,1] ~~{\rm and}
&& \!\!\!\!
\xi_{\nu' \nu}
= \xi_{\nu \nu'},
\cr
t {\rm ~channel:~~}~
\tau _{\nu_1 \nu_2 | \nu_3 \nu_4}
&\equiv&
\exp[\pi i(\zeta_{\nu_1 \nu_4} + \zeta_{\nu_2 \nu_3})];
\cr
~~~ ~~~ ~~~ 
\zeta_{\nu\nu'} \in [-1,1] ~~{\rm and}
&& \!\!\!
~\zeta_{\nu'\nu} = -\zeta_{\nu\nu'},
\cr
u {\rm ~channel:~~}
\upsilon _{\nu_1 \nu_2 | \nu_3 \nu_4}
&\equiv&
\exp[\pi i(\vartheta_{\nu_1 \nu_3} + \vartheta_{\nu_2 \nu_4})];
~~~ ~~~ 
\cr
~~~ ~~~ ~~~ 
\vartheta_{\nu \nu'} \in [-1,1] ~~{\rm and}
&& \!\!\!\!
\vartheta_{\nu' \nu} = -\vartheta_{\nu \nu'}.   
\label{kII12}
\end{eqnarray}
Their full outworkings are detailed in KI.
The uniformly random numbers $\xi, \zeta, \vartheta$ are independently
distributed; expectations over them mutually decouple and all factors
conform to Eq. (\ref{kII10}). Each is designed so that, in
the stochastic average of the diagrammatic expansion of $\Phi$, product
chains whose phases cancel identically from start to finish are
immune to the averaging. All other product chains fail to cancel.
Being stochastic they interfere destructively, vanishing in the limit of
an arbitrarily large ensemble.

With respect to $t$ and $u$ channels, note that an exchange
of labels $1 \leftrightarrow 2$ or
$3 \leftrightarrow 4$ effectively swaps the definitions
and thus the actions of their K-couplings. This is consistent
with the physics of these channels as mutual exchange counterparts.

In the modality of Eq. (\ref{kII12}), $\sigma $ generates the
particle-particle Brueckner-ladder functional while the
ring approximation is generated by $\tau $.
Last, $\upsilon $ also results in a
Brueckner-like functional where particle-hole ladders replace
particle-particle ones.
\cite{KI}
There are no other options for pairwise propagation,
just as with parquet.

\subsection{Maximal pairwise coupling}

None of the K-couplings of Eq. (\ref{kII12}), alone or in twos, can
cover all conceivable scattering arrangements strictly
between particle and/or hole propagator pairs.
All three must combine sequentially in all possible ways,
while preventing any potential
replication of terms if two or more K-couplings led to the survival of
identical $\Phi$ terms. To first and second order in
$\OV$ one can show that all three elementary couplings
generate identical contributions, inducing overcounting which would
propagate throughout the nesting of self-energy insertions.

The solution to overcounting is to combine the couplings of
Eq. (\ref{kII12}) to inhibit any concurrency. We propose the
candidate parquet K-coupling to be
\begin{eqnarray}
\varphi
&\equiv&
1 - (1 - \sigma )(1 - \tau )(1 - \upsilon ) ~~{\rm so}
\cr
\varphi_{\nu_1 \nu_2 | \nu_3 \nu_4}   
&=&
\sigma _{\nu_1 \nu_2 | \nu_3 \nu_4} + \tau _{\nu_1 \nu_2 | \nu_3 \nu_4}
+ \upsilon _{\nu_1 \nu_2 | \nu_3 \nu_4}
\cr
&&
-~ \sigma _{\nu_1 \nu_2 | \nu_3 \nu_4} \tau _{\nu_1 \nu_2 | \nu_3 \nu_4}
\cr
&&
-~ \tau _{\nu_1 \nu_2 | \nu_3 \nu_4} \upsilon _{\nu_1 \nu_2 | \nu_3 \nu_4}
\cr
&&
-~ \upsilon _{\nu_1 \nu_2 | \nu_3 \nu_4} \sigma _{\nu_1 \nu_2 | \nu_3 \nu_4}
\cr
&&
+~ \sigma _{\nu_1 \nu_2 | \nu_3 \nu_4} \tau _{\nu_1 \nu_2 | \nu_3 \nu_4}
\upsilon _{\nu_1 \nu_2 | \nu_3 \nu_4},
\label{kII13}
\end{eqnarray}
preserving overall the Hermitian property specified by Eq. (\ref{kII10}).

In any diagram expanded to a given order in $\OV$,
the products of K-couplings in Eq. (\ref{kII13}) may or may not
resolve into a set of elementary closed cycles
whose multiplicative chain is identically unity when $\varphi$-averaged
(this means, by way of definition, that any sub-chain hived off
within an elementary cycle would necessarily vanish through
phase interference).
Chains not resolving into a set of independent closed cycles
over the contribution will be quenched to vanish
in the Kraichnan expectation.

In Fig. 4(a) we show schematically the structures of the three
possible pairwise multiple-scattering combinations contributing
explicitly to the correlation energy functional $\Phi$, Fig. 4(b).
The K-coupling $\sigma $ leads to the particle-particle Brueckner
t-matrix $\Lambda_s$, while $\tau $ leads to the screened interaction
$\Lambda_t$ and lastly $\upsilon $ is the $t$-exchange complement leading
to the particle-hole Brueckner-type ladder $\Lambda_u$; this carries
an implicit sign change relative to $\Lambda_t$ owing to the difference
of one fermion loop count. From Eq. (\ref{kII13}) all processes
combine so the renormalized one-body propagators $G$ carry
self-energy insertions to
all orders in which $s$-, $t$- and $u-$ processes act synergetically, not
competing in parallel but entering sequentially.

\centerline{
\includegraphics[height=4.5truecm]{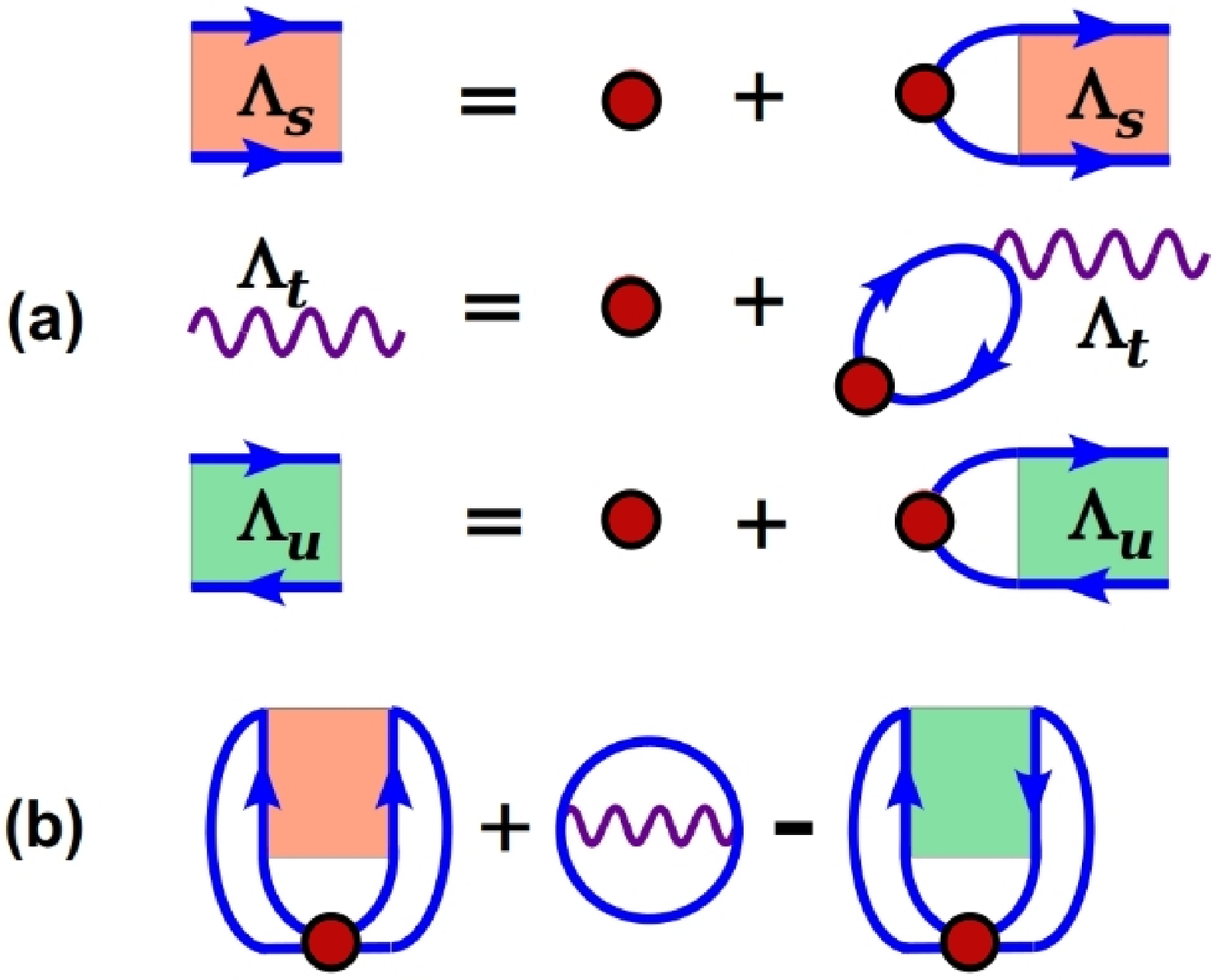}
}
\vskip 0.25cm
{{\bf FIG. 4.} {\small (a) Definition of the fundamental all-order
$s, t$ and $u$ interactions. Dots: antisymmetrized pair potential.
(b) Symbolic definition of $\Phi$, the correlation energy functional
(weightings induced by Eq. (\ref{kII02}) are understood), following Kraichnan
averaging over all K-couplings $\sigma , \tau , \upsilon $ as in
Eq. (\ref{kII13}) to remove overcounting when different K-couplings
lead to identical diagrams. Although the skeleton graphs for $\Phi$
appear simple, their complexity is hidden within the self-consistent
nesting of self-energy insertions in the propagators (solid lines)
according to Eqs. (\ref{kII02})--(\ref{kII04}). Since the $stu$
correlation energy is identical to that of the fluctuation-exchange model
\cite{pqt3,sb},
the Kraichnan construction already subsumes the essence of parquet.
The combinatorial $stu$ structure is fully revealed only when the
response to an external perturbation is extracted
(see following).
}}
\vskip 0.2cm

Should two or even three channels have coincident closed cycles,
the structure of $\varphi$ makes certain that the net contribution
from this coincidence is always precisely unity;
Eq. (\ref{kII13}) ensures
that there is no overcounting if $GG$ pairings from different
channels gave rise to the same diagrammatic structure.
Such terms can turn up only once in their locations within
the expansion for $\Phi$, including iteratively in the self-energy parts.

All allowed pairwise-only combinations of scatterings, and only those,
survive the expectation over $\{\varphi\}$ to lead to a legitimate
$\Phi$-derivable correlation term with all its 
symmetries and conserving properties.
\cite{k1,KI,kb2}
The individual energy
functional of each component Hamiltonian in Eq. (\ref{grh}),
being exact in its particular configuration prior to averaging,
automatically has these symmetries
in the renormalized expansion of Eq. (\ref{kII02}).
These are inherited
by the diagrammatic structure of every term that
survives the taking of expectations and ultimately by
the complete averaged $\Phi$.\\

\subsection{$\Phi$-derivable response}

Having arrived at the maximally paired structure of $\Phi$ in Fig. 4(b)
given the K-couplings of Eq. (\ref{kII13}), the work of obtaining the
parquet equations from it has been done, in one sense, in the analysis
detailed by Bickers
\cite{pqt3}
for the equivalent heuristic FLEX model.
However, the Hamiltonian prescription's ramifications lead beyond
the derivation of classic parquet.

\centerline{
\includegraphics[height=7.0truecm]{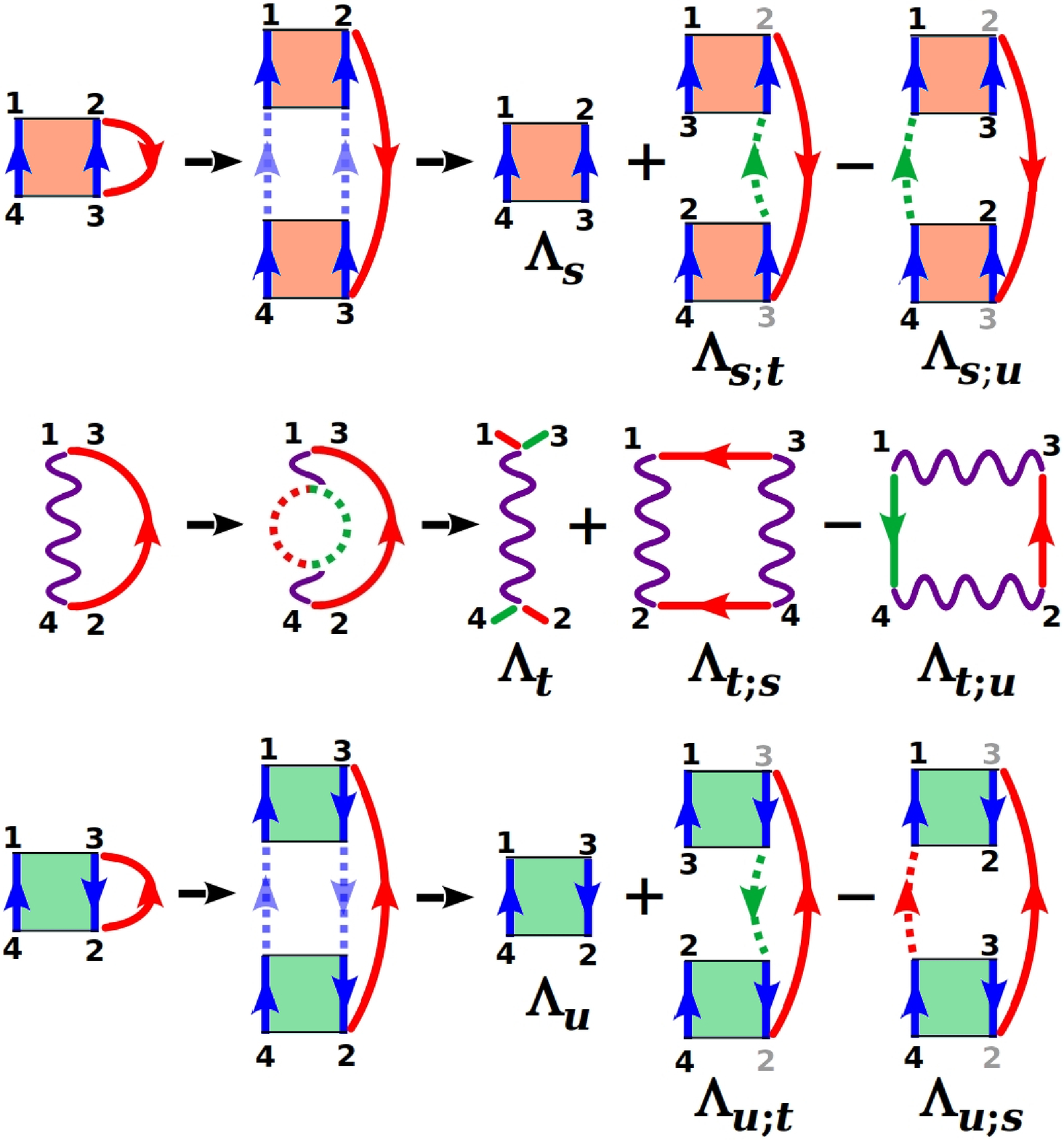}
}
{{\bf FIG. 5.} {\small Systematic removal
of a propagator $G$ internal to the self-energy
$\Sigma[\varphi\overline V; G] = \Lambda\!:\!G$,
after Baym and Kadanoff,
\cite{kb1,kb2}
generates the primitive scattering kernel $\Lambda'$.
Removal of $G(32)$, solid line, simply regenerates $\Lambda$.
Removing any internal $G$ other than $G(32)$ yields additional terms
required for $\Phi$ derivability (microscopic conservation).
Top line: beyond the $s$-channel ladder $\Lambda_s$ the
non-crossing symmetric $t$-like term $\Lambda_{s;t}$ and
$u$-term $\Lambda_{s;u}$ are generated.
Middle line: generation of $\Lambda_t$ and the non-symmetric
$\Lambda_{t;s}$ and $\Lambda_{t;u}$. Bottom line: generation of
$\Lambda_u$ with $\Lambda_{u;t}$ and $\Lambda_{u;s}$.
}}
\vskip 0.20cm

The goal, then, is to reconstruct
a parquet-like scattering amplitude $\Gamma$ using the ingredients
provided by the Kraichnan machinery. We are still left to show
its relation to the scattering function $\Lambda$,
generator of the $\Phi$-derivable diagrammatic $stu$ expansion.
While the renormalized structure of $\Lambda$ seems sparse
compared with $\Gamma$ for parquet,
\cite{pqt3}
the structure for actual comparison is not $\Lambda$ but begins
with the variation
\begin{eqnarray}
\Lambda'
&\equiv&
\frac{\delta \Sigma}{\delta G} = \frac{\delta^2 \Phi}{\delta G \delta G},
\label{kII18}
\end{eqnarray}
which in fact is the source of the Ward-Pitaevsky identities.
\cite{pn}
One goes from there to set up the complete scattering interaction
$\Gamma'$ for the total system response to a perturbation.

The full outcome of the derivation of $\Gamma'$ is the
dynamical theory of Baym and Kadanoff;
\cite{kb2}
in it, conservation entails the additional family of non-parquet
diagrams $\Lambda'' = \Lambda' - \Lambda$ shown in Fig. 5.
These contribute to every order of iteration.
The topologies contained in $\Lambda''$ are {\em not} explicit in
the renormalized $\Lambda$ embedded within $\Phi[\varphi\OV;G]$.
Not being crossing symmetric they are not permitted, much less generated,
within parquet. As with the normal parquet structures that we aim to
exhibit from the stochastic Hamiltonian construction, the apparently
extra correlation effects, actually mandated by conservation, remain
virtual in the renormalized summation for $\Phi$ until elicited by an
external probe.

Figure 5 details how functional differentiation gives rise to the
non-parquet terms, typical of all $\Phi$-derivable descriptions.
We want to trace how the purely parquet crossing symmetric
$\Gamma$ diagrams make up a nontrivial component of the complete set for
$\Gamma'$, the total Baym-Kadanoff response kernel. $\Gamma$
is not equivalent to $\Gamma'$; it is a proper subset.
\cite{pqt3}

We emphasize the necessary presence, for $\Phi$ derivability, of the
non-symmetric components $\Lambda''$. These are the approximate system's
attempt to match its $u$ terms, for example,
with partner terms topologically like the two complementary
channels $t$ and $s$; the same applies correspondingly to
the primary $s$ and $t$ terms. However, the question is less why
they break antisymmetry but how their presence fits into the
cancellation of terms for conservation to govern the model's response.

\section{Derivation of the parquet equations}

\subsection{Origin within response analysis}

To unpack the nested correlations hidden in
the renormalized form of $\Phi$
we turn to the full Kraichnan Hamiltonian prior to averaging
and derive the response to a one-body nonlocal perturbation
${\langle k' | U | k \rangle}$, which generally will have a
time dependence also.
\cite{kb1,kb2}
External perturbations do not couple
to the collective index $\nu$ but physically only to labels
$k$. The interaction Hamiltonian in Eq. (\ref{kII11}) is augmented:
\begin{eqnarray}
{\cal H}_{i;N}[\varphi; U]
&\equiv&
\sum_{ll'} {\langle k' | U | k \rangle} a^*_{l'} a_l
+ {\cal H}_{i;N}[\varphi; U=0].
~~~ 
\label{kII18.1}
\end{eqnarray}
Response to a local field is generated by setting
${\langle k' | U | k \rangle} \to U(q)\delta_{k',k+q}$,
dynamically linking (contracting)
the propagators that terminate and start at $U$.

Physical expectations are taken next, while
retaining the individual K-couplings $\varphi$
to keep track of all pair processes.
We use matrix notation with repeated indices
to expand the intermediate sums.

The two-body Green function is $\delta G/\delta U$.
\cite{kb1}
Working from Eq. (\ref{kII03}), vary $G^{-1}$ for
\begin{widetext}
\begin{eqnarray*}
\delta G^{-1}(12)
&=&
- \delta U(12) - \delta \Sigma[\varphi,G](12)
~~~{\rm or}
\cr
G^{-1}(12') \delta G(2'1') G^{-1}(1'2)
&=&
\delta U(12) + \frac{\delta \Sigma(12)}{\delta G(43)}
\frac{\delta G(43)}{\delta U(56)}
~~{\rm so}
\cr
\frac{\delta G(21)}{\delta U(56)}
\equiv
G(25)G(61)
&&\!\!\!\!\!\!
+~ G(21')G(2'1) \Lambda'(1'3|2'4)\varphi_{1'3|2'4}
\frac{\delta G(43)}{\delta U(56)}
\end{eqnarray*}
where $\varphi$ explicitly partners the
effective interaction $\Lambda'$. There is no overcounting of
the primary $s,t$ and $u$ contributions of $\Lambda$ since,
once a line in any self-energy insertion is opened, it will not
reconnect to its originating structure, joining instead a new
and different (ultimately closed) loop. Symbolically, with $I$
the two-point identity,
\begin{eqnarray}
[II - GG\!:\!\Lambda'\varphi]\!:\!\frac{\delta G}{\delta U}
&=&
GG ~~{\rm so}
\cr
\cr
\frac{\delta G}{\delta U}
&=&
[II - GG\!:\!\Lambda'\varphi]^{-1}\!:\!GG
\cr
\cr
&=&
GG + GG\!:\![II - GG\!:\!\Lambda'\varphi]^{-1}\Lambda'\varphi\!:\!GG.
\label{kII18.3}
\end{eqnarray}
\end{widetext}
Recalling Eq. (\ref{kII02}), the form of the generating kernel
$\Lambda$ (without the non-crossing-symmetric components 
from Fig. 5) can be read off from the structure of $\Phi$
as in Fig. 4, with the subsidiary kernels
$\Lambda_s, \Lambda_t$ and $\Lambda_u$:
\begin{eqnarray}
\Lambda
&=&
\Lambda_s + \Lambda_t - \Lambda_u
~~ {\rm where}
\cr
\cr
\Lambda_s
&=&
\OV + {\phi}^{-1}\OV\sigma  \!:\! GG\!:\!\Lambda_s\varphi;
\cr
\Lambda_t
&=&
\OV + {\phi}^{-1}\OV\tau    \!:\! GG\!:\!\Lambda_t\varphi;
\cr
\Lambda_u
&=&
\OV + {\phi}^{-1}\OV\upsilon\!:\! GG\!:\!\Lambda_u\varphi.
\label{kII15}
\end{eqnarray}
To put the interactions on the same
representational footing as $\OV$, we factor out the outermost
K-coupling, ${\phi}$. Intermediate chains that cancel right across
will finally cancel with ${\phi}^{-1}$ appropriate to each channel.
In Eq, (\ref{kII15}) the $u$-channel term of $\Lambda$,
being the exchange of the $t$-channel, carries the sign
tracking the structural antisymmetry of $\Lambda_t$
on swapping particle (or hole) end points and restoring to $V$
its proper weight of unity in the intermediate summations.

From Eq. (\ref{kII18.3}) the complete four-point amplitude is defined:
\begin{eqnarray}
\Gamma'
&\equiv&
{\phi}^{-1}\Lambda'\varphi\!:\![II - GG\!:\!\Lambda'\varphi]^{-1}
\cr
&=&
{\phi}^{-1}[II - \Lambda'\varphi\!:\!GG]^{-1}\!:\!\Lambda'\varphi\!,
\label{kII18.4}
\end{eqnarray}
In terms of $\Gamma'$ the conserving two-body Green function becomes
\begin{eqnarray}
\frac{\delta G}{\delta U}
&=&
GG\!:\![II + \Gamma'\varphi\!:\!GG].
\label{kII18.5}
\end{eqnarray}

\subsection{Parquet equations}

At this stage we specialize to the crossing symmetric sub-class of the
expansion dictated by Eq. (\ref{kII18.4}).
After dropping $\Lambda''$ all the crossing-symmetric terms
are gathered. The Equation is truncated and defines the
now crossing symmetric kernel
\begin{eqnarray}
\Gamma
&\equiv&
 {\phi}^{-1}\Lambda\varphi\!:\![II - GG\!:\!\Lambda\varphi]^{-1}
\cr
&=&
 {\phi}^{-1}[II - \Lambda\varphi\!:\!GG]^{-1}\!:\!\Lambda\varphi\!,
\label{kII18.6}
\end{eqnarray}
keeping in mind that the crossing symmetric $\Lambda$ consists
only of the primary structures embedded in $\Phi$. While $\Gamma$
inherits antisymmetry, it forfeits conservation at the two-body level
that is guaranteed for $\Gamma'$.
\cite{err}

As with Eq. (\ref{kII18.4}) above,
Eq. (\ref{kII18.6}) sums $\Gamma$ differently from parquet;
but the underlying architecture of $\Gamma$ is the same.
In the Equation, $s$-, $t$- and $u$-processes combine
in all possible ways while inhibited from acting concurrently.
The resolution of $\Gamma$ becomes a bookkeeping exercise:
to make a systematic species by species inventory of all its
permissible pair-only scattering sequences, irreducible
in the parquet sense, within each channel, finally to weave these
into all possible reducible contributions.

In Kraichnan's description one resums $\Gamma$ by tracking how selective
filtering works through the three possible K-couplings, while in the
parquet approach one enforces, on the intermediate $GG$ pairs, the three
distinct modes of momentum, energy and spin transfer characterizing
the $s, t$ and $u$ channels. Our operation is the same as the pairwise
topological argument for $\Gamma$ in FLEX, detailed in Ref.
\cite{pqt3}.
To achieve it, the first set of equations isolates the components
that are not further reducible within each particular channel:
\begin{eqnarray}
\Gamma_s
&\equiv&
\OV
+ {\phi}^{-1} \Gamma\tau    \!:\! GG\!:\!\Gamma_t\varphi
- {\phi}^{-1} \Gamma\upsilon\!:\! GG\!:\!\Gamma_u\varphi;
\cr
\Gamma_t
&\equiv&
\OV
- {\phi}^{-1} \Gamma\upsilon\!:\! GG\!:\!\Gamma_u \varphi
+ {\phi}^{-1} \Gamma\sigma  \!:\! GG\!:\!\Gamma_s\varphi;
\cr
\Gamma_u
&\equiv&
\OV
+ {\phi}^{-1} \Gamma\sigma  \!:\! GG\!:\!\Gamma_s\varphi 
+ {\phi}^{-1} \Gamma\tau    \!:\! GG\!:\!\Gamma_t\varphi.
~~~ ~~~ 
\label{kII15.2}
\end{eqnarray}
Manifestly, the components of $\Gamma_s$ couple only via $t$ or $u$,
excluding any $s$-channel processes where cutting a pair sequence
$\sigma GG$ yields two detached diagrams. Thus, taking the Kraichnan
expectation of $(\Gamma_s - \OV)\sigma$, nothing survives; and so on
for the other channels. The arrangement generates every legitimate
convolution involving internally closed cycles of propagation through
every channel within $\Gamma$ while ensuring irreducibility of the
three component kernels.

Finally the complete $\Gamma$ is assembled:
\begin{eqnarray}
\Gamma
&=&
\OV
+ {\phi}^{-1} \Gamma\sigma  \!:\! GG\!:\!\Gamma_s\varphi
+ {\phi}^{-1} \Gamma\tau    \!:\! GG\!:\!\Gamma_t\varphi
\cr
&&
~~~ ~~~ 
- {\phi}^{-1} \Gamma\upsilon\!:\! GG\!:\!\Gamma_u\varphi.
\label{kII15.3}
\end{eqnarray}
In the Kraichnan average only the pairwise terms we have highlighted
will survive. Then Eqs. (\ref{kII15.2}) and (\ref{kII15.3}) become
identical to the FLEX parquet equations.
\cite{pqt3}
For each channel the total amplitude can also be recast
to show its reducibility explicitly:
\begin{eqnarray*}
\Gamma
&=&
\Gamma_s + {\phi}^{-1} \Gamma\sigma  \!:\!GG\!:\!\Gamma_s\varphi
\cr
&=&
\Gamma_t + {\phi}^{-1} \Gamma\tau    \!:\!GG\!:\!\Gamma_t\varphi
\cr
&=&
\Gamma_u - {\phi}^{-1} \Gamma\upsilon\!:\!GG\!:\!\Gamma_u\varphi.
\end{eqnarray*}

\subsection{Extension of the parquet equations}

\subsubsection{Complete specification of $\Gamma'$}

The $stu$ based formalism leads to an
interaction energy functional $\Phi$ equivalent to the
fluctuation-exchange approximation introduced by Bickers {\em et al.}.
\cite{sb}
The radical difference is that its properties are not imparted
intuitively; they are established from a Hamiltonian. The
consequence of this canonical provenance is to set a limit
on what is possible diagrammatically for a conserving,
optimally pairwise-correlated model.

The FLEX model generates the parquet topology naturally by generating,
as we have done within Kraichnan's formalism, the variational
structure of the $\Phi$-derivable self-energy $\Sigma$.
Without the additional step of obtaining the perturbative response,
the intimate link between the renormalized topology of $\Phi$
and the architecture of parquet, which is otherwise implicit within
the self-consistent correlation energy functional, does not emerge.

A widespread line of thought in parquet literature assumes there is
no distinction between, on the one hand, the scattering kernel $\Lambda$
in the self-energy $\Sigma$ and, on the other,
$\Gamma$ acting as the kernel for the total two-body response
within parquet. For $\Phi$-derivable models this is not permissible,
if only because they yield two different numerical
estimates for the static pair correlation function,
of which only one meets the exact formulation
by functional differentiation of $\Phi$ with respect to $V$.
\cite{KI,fgetal}
This reflects the loss of Fock-space completeness.
Later we revisit the implications, for consistency in conservation,
of the parquet model's assumption $\Lambda \equiv \Gamma$.

We have a basis to build up more elaborate extensions of the FLEX parquet
model following the Kraichnan-based analysis. First, the complete
$\Phi$-derivable, conserving pair-scattering kernel $\Gamma'$
may always be obtained from the original representation
Eq. (\ref{kII18.4}), but it is more practical to adapt the parquet
equations (\ref{kII15.2}) and (\ref{kII15.3}). Noting that the
non-parquet term $\Lambda'' = \Lambda' - \Lambda$ is absolutely
irreducible within both parquet and Kadanoff-Baym (recall Fig. 5), 
the bare potential is replaced with
\[
{\cal V} \equiv \OV + \Lambda''.
\]

Return to Eq. (\ref{kII15.2}), this time to define
\begin{eqnarray}
\Gamma'_s
&\equiv&
{\cal V}
+ {\phi}^{-1} \Gamma'\tau    \!:\!GG\!:\!\Gamma'_t\varphi
- {\phi}^{-1} \Gamma'\upsilon\!:\!GG\!:\!\Gamma'_u\varphi;
\cr
\Gamma'_t
&\equiv&
{\cal V}
- {\phi}^{-1} \Gamma'\upsilon\!:\!GG\!:\!\Gamma'_u\varphi 
+ {\phi}^{-1} \Gamma'\sigma  \!:\!GG\!:\!\Gamma'_s\varphi;
\cr
\Gamma'_u
&\equiv&
{\cal V}
+ {\phi}^{-1} \Gamma'\sigma  \!:\!GG\!:\!\Gamma'_s\varphi
+ {\phi}^{-1} \Gamma'\tau    \!:\!GG\!:\!\Gamma'_t\varphi
~~~ ~~~ 
\label{kII15.5}
\end{eqnarray}
with the ultimate result
\begin{eqnarray}
\Gamma'
&=&
{\cal V}
+ {\phi}^{-1} \Gamma'\sigma  \!:\!GG\!:\!\Gamma'_s\varphi
+ {\phi}^{-1} \Gamma'\tau    \!:\!GG\!:\!\Gamma'_t\varphi
\cr
&&
~~~ 
- {\phi}^{-1} \Gamma'\upsilon\!:\!GG\!:\!\Gamma'_u\varphi.
\label{kII15.6}
\end{eqnarray}
We stress that the only feature that matters in
Eqs. (\ref{kII15.5}) and (\ref{kII15.6})
is the topological arrangement of the elements of the response kernel,
exhausting all possible interplays among the three $\Phi$-derivable
channels independently of crossing symmetry.

Now we address the addition to the energy functional, presumably by
physical intuition, of absolutely pair-irreducible graphs for $\Phi$.
(Two are shown in Fig. 3.)

\subsubsection{Contributions from pair-irreducible terms of $\Phi$}

\centerline{
\includegraphics[height=2.truecm]{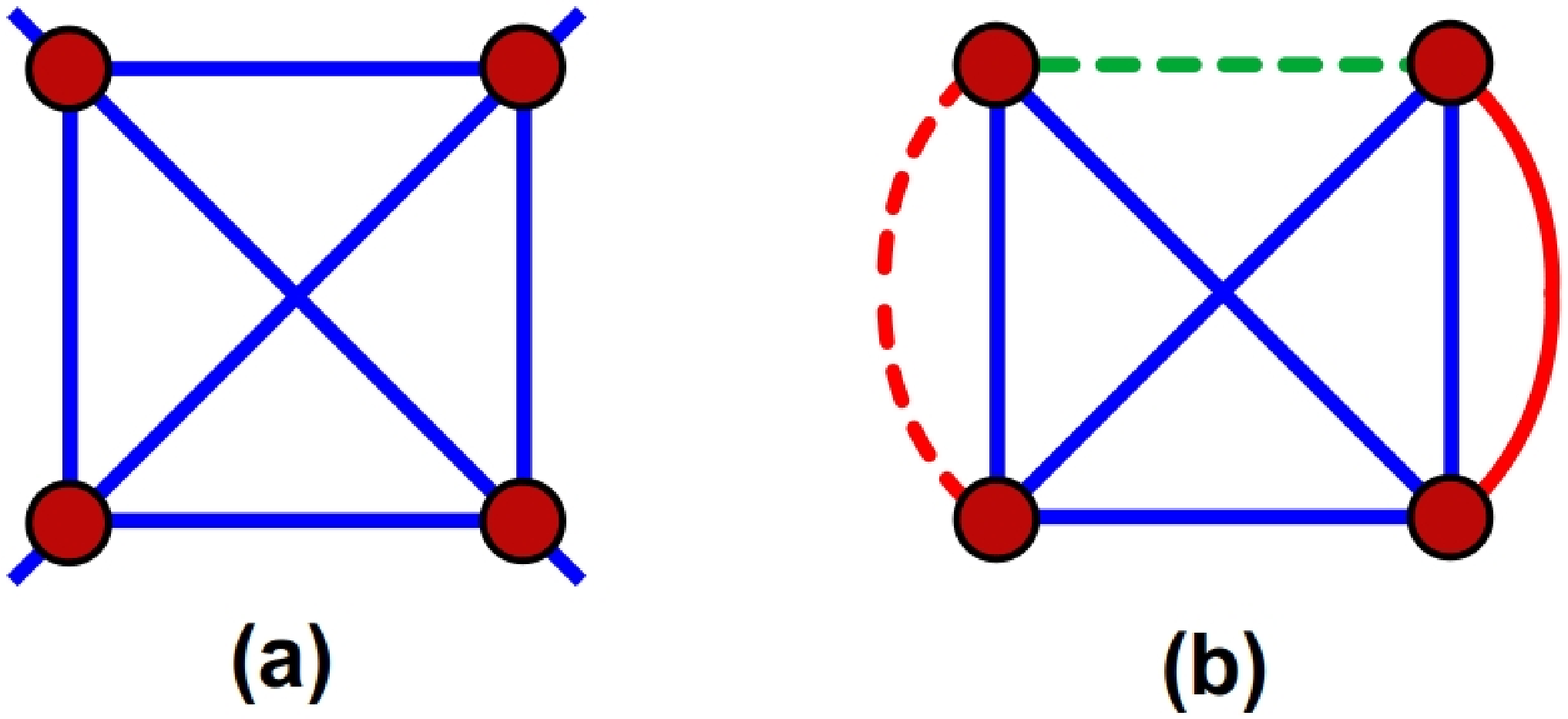}
}
{{\bf FIG. 6.} {\small (a) Fourth-order $stu$-irreducible crossing symmetric
graph, valid as a primitive input to the standard parquet equations but
not $\Phi$-derivable. While the graph can be generated by removing an
interaction node from its analog in Fig. 3(a), when closed with two final
propagators as in (b) it is forced to carry inequivalent propagators
(dotted lines). Thus it is disqualified from any $\Phi$-derivable
approximation since it cannot lead to a unique self-energy functional.
}}
\vskip 0.20cm

Primitive additions to $\Lambda'$ can be incorporated once again via
Eqs. (\ref{kII15.5}) and (\ref{kII15.6}). 
In $\Phi$ derivability the choice of symmetric structures for the
$\Lambda$ kernel is highly constraining. Readers can convince themselves,
with a bit of sketching, that no such three-node term exists.
Nor is there an $stu$-irreducible four-node term for
$\Phi$ that has the needed symmetry. Whereas the crossing symmetric
four-node ``envelope'' graph depicted in Fig. 6(a) is a valid
irreducible interaction in parquet,
\cite{pqt3}
when incorporated as a fully closed diagram it must carry
inequivalent propagators, making it inadmissible in any
$\Phi$-derivable subset of the correlation energy.

The next-order $\Phi$-derivable skeleton beyond second 
is that of Fig. 3(a), with five interaction nodes.
Its variation with respect to any node -- removal
of a node from Fig. 3(a) -- generates a two-body correlation
with parquet's envelope graph as its kernel.
However there is no systematic link between such a variation and parquet.

The issue with adding higher-order $stu$-irreducible terms to $\Phi$,
again assuming that they held some novel physical effects,
is that one gets back to adding many-body correlations heuristically,
without a Hamiltonian basis. Strictly, then, the sum-rule identities
no longer come for free but require individual validation
(this has been done up to the third-frequency-moment rule
\cite{fgetal}).
Kraichnan's procedure is limited to pair interactions; so far, it is hard
to envisage how any Hamiltonian extension could generate these
additional complex objects. Nevertheless adding a totally
pairwise-irreducible structure satisfying Baym-Kadanoff symmetry will
not spoil $\Phi$ derivability.

\subsection{Parquet and $\Phi$-derivability}

In establishing full parquet the $\Phi$-derivable FLEX approximation
has been taken as a suitable entry point for successive iterations aimed
at approaching the full structure;
but the initial self-energy $\Lambda\!:\!G$ is considered to
fall short of a maximally correlated parquet. It is deemed
necessary to feed the FLEX-derived crossing symmetric $\Gamma$ in
Eq. (\ref{kII15.3}) back into $\Sigma$ in Eq. (\ref{kII04}) via the
replacements
\cite{pqt3}
(ensuring that $G\!\!:\!\Gamma\!\!:\!G$ does not double
up on terms previously included)
\begin{eqnarray}
\Sigma(13)
&\leftarrow& 
{\widehat \Gamma}(12|34)G(42) ~~\text{in which}~~
\cr
\cr
{\widehat \Gamma}(12|34)
&\leftarrow&
\OV(12|34)
\cr
&&
+~ \Gamma(12|3'4')G(4'2')G(3'1')\OV(1'2'|34),
~~~ ~~ 
\label{lw47.0}
\end{eqnarray}
with the nonconforming piece, $\Gamma'' = \Gamma' - \Gamma$, naturally
absent. Substitution of ${\widehat \Gamma}$ for $\Lambda$ in the self-energy
assumes that no distinction should be made between
the approximate self-energy kernel and the approximate two-body response
kernel: that, as in the exact theory, they are one and the same.
\cite{pqt2a,pqt3}
As a generator of new primitively irreducible structures
Eq. (\ref{lw47.0}) can be iterated at will.

In view of how the generic parquet Equations (\ref{kII15.5})
and (\ref{kII15.6}) always build up from at least the leading primitive
irreducible, namely $\OV$, any resulting $\Gamma$ must always incorporate
$\Lambda$ from Eq. (\ref{kII15}). Reopening lines in
the self-energy ${\widehat \Gamma}[V, G]\!:\!G$ is always going to regenerate
pieces including the unwanted non-parquet term $\Lambda''$.

To compare the behaviors of the different self-energy kernels for
$stu$ and standard parquet, we use a result of Luttinger and Ward.
\cite{lw}
Equation (47) of that Reference
provides an alternative formulation of the correlation energy when
$\Lambda$ in Eq. (\ref{kII02}) is the exact $\Gamma$ interaction:
\begin{eqnarray}
\Phi[V;G]
&=&
- {\langle \ln(I - G^{(0)}\!\cdot\!\Sigma~\cdot) \rangle}
- G[V]\!:\!\Sigma
\cr
&&
+~ \int^1_0 \frac{dz}{2z} G[V]\!:\!\Gamma[zV;G[V]]\!:\!G[V].
\label{lw47.1}
\end{eqnarray}
The difference between the coupling-constant integral on the right-hand
side of this identity and its counterpart in Eq. (\ref{kII02}) is that the
former keeps track only of the combinatorial factors for the $V$s in the
original linked skeleton diagrams $\Gamma[V;G^{(0)}]$ but now with $G[V]$,
containing $V$ at full strength, in place of each bare line $G^{(0)}$.
By contrast, in the integral on the right-hand side of Eq. (\ref{kII02})
the coupling factor attaches to all occurrences of $V$; that is,
including those within $G[V]$ itself.

The correlation energy as given in Eq. (\ref{lw47.1}) leads to two
identities. Exploiting the equivalence of all propagators in the closed
structure $G\!\!:\!\Gamma\!\!:\!G$ within the integral,
varying on both sides with respect to the self-energy gives
\begin{eqnarray}
\frac{\delta \Phi}{\delta \Sigma}
&=&
(I - G^{(0)}\!\cdot\!\Sigma)^{-1}\!\cdot\!G^{(0)} - G
- \frac{\delta G}{\delta \Sigma}\!:\!\Sigma
\cr
&&
~~~ ~~~ 
+ \frac{\delta G}{\delta \Sigma}\!:\!\Gamma[V; G]\!:\!G
\cr
&=&
- \frac{\delta G}{\delta \Sigma}\!:\!(\Sigma - \Gamma[V; G]\!:\!G)
\cr
&=&
0
\label{lw47.2}
\end{eqnarray}
on using Eq. (\ref{kII04}).
The vanishing of this derivative establishes the
correlation energy as an extremum with respect to perturbations,
as these add linearly to $\Sigma$.

Next,
\begin{eqnarray}
\frac{\delta \Phi}{\delta G}
&=&
{\Bigl( (I - G^{(0)}\!\cdot\!\Sigma)^{-1}\!\cdot\!G^{(0)} - G \Bigr)}
\!\cdot\!\frac{\delta \Sigma}{\delta G}
\cr
&&
~~~ ~~~ 
+~ \Gamma[V; G]\!:\!G
\cr
&=&
\Sigma.
\label{lw47.3}
\end{eqnarray}
Consistency with Eq. (\ref{kII04}) is confirmed.

We look at how Eq. (\ref{lw47.1}) works in the $\Phi$-derivable case.
Since it applies canonically in the case of the full Kraichnan Hamiltonian,
the form survives the expectation over the $stu$ couplings, as will the
form of the variational derivatives; the skeletal topology of the
integrals on the right sides of Eqs. (\ref{kII02}) and (\ref{lw47.1})
is the same. In the
stochastic expectations on the right-hand side of
Eq. (\ref{lw47.1}), $\Gamma$ goes over to the reduced $stu$ structure
$\Lambda$ depicted in Fig. 4. This is because, in the coupling-constant
integral, the pattern of surviving and suppressed products of factors
$\varphi$ is identical with that leading to Eq. (\ref{kII02}).

Define the $\Phi$-derivable correlation energy $\Phi_{\rm KB}$ from the
corresponding Eq. (\ref{lw47.1}).
All propagators in the structure $G\!\!:\!\Lambda\!\!:\!G$ being
equivalent, the variation in Eq. (\ref{lw47.2}) again leads to
\begin{eqnarray}
\frac{\delta \Phi_{\rm KB}}{\delta \Sigma}
&=&
- \frac{\delta G}{\delta \Sigma}\!:\!(\Sigma - \Lambda\!:\!G)
= 0
\label{lw47.4}
\end{eqnarray}
so the extremum property holds for the approximate correlation energy.
The relation Eq, (\ref{lw47.3}) becomes
\begin{eqnarray}
\frac{\delta \Phi_{\rm KB}}{\delta G}
&=&
\Lambda\!:\!G = \Sigma
\label{lw47.5}
\end{eqnarray}
since the symmetry of the integral $G\!:\!{\delta\Lambda/\delta G}\!:\!G$
works once more as for Eq. (\ref{kII04}) to recover the self-energy.

The analysis is now applied to the classic parquet expansion,
whose candidate correlation energy functional,
defined from Eq. (\ref{lw47.1}),
we will call $\Phi_{\rm PQ}$. In this instance one gets
\begin{eqnarray}
\frac{\delta \Phi_{\rm PQ}}{\delta \Sigma}
&=&
- \frac{\delta G}{\delta \Sigma}\!:\!{\Bigl( \Sigma - \Gamma[V; G]\!:\!G
- \Delta\Gamma[V; G]\!:\!G \Bigr)};
~~~ 
\cr
\cr
\Delta\Gamma[V; G]
&\equiv&
\int^1_0 \frac{dz}{z} ( \Gamma[zV;G[V]] - z\Gamma[V;G[V]] )
\cr
&&
- \int^1_0 \frac{dz}{2z}
\frac{\delta \Gamma[zV;G[V]]}{\delta G}\!:\!G[V].
\label{lw47.6}
\end{eqnarray}
This does not vanish because the parquet structure $G\!:\!\Gamma\!:\!G$
contains inequivalent propagators. Therefore Eq. (\ref{lw47.2}) fails.
The same  topological absence of $\Phi$-derivable symmetry spoils the
complementary attempt to define $\Phi_{\rm PQ}$ from the alternative
fundamental expression Eq. (\ref{kII02}).

Figure 7 typifies the issue. At third order in the bare potential
the parquet iteration of $\Gamma$ obtained from FLEX produces a new
absolutely irreducible term at fourth order whose skeleton
contribution to $\Phi$ would carry inequivalent propagators,
as already shown in Fig. 6(b).
Since the symmetry leading to a well defined $\Phi$ must be present
at all orders it follows that no formulation of parquet, built on
pairwise-only scattering, can be $\Phi$-derivable.
Conversely, the $stu$ construction \`a la Kraichnan is the only strictly
pairwise-correlated model that has a Hamiltonian basis while exhibiting
the essential parquet topology.

\centerline{
\includegraphics[height=4.0truecm]{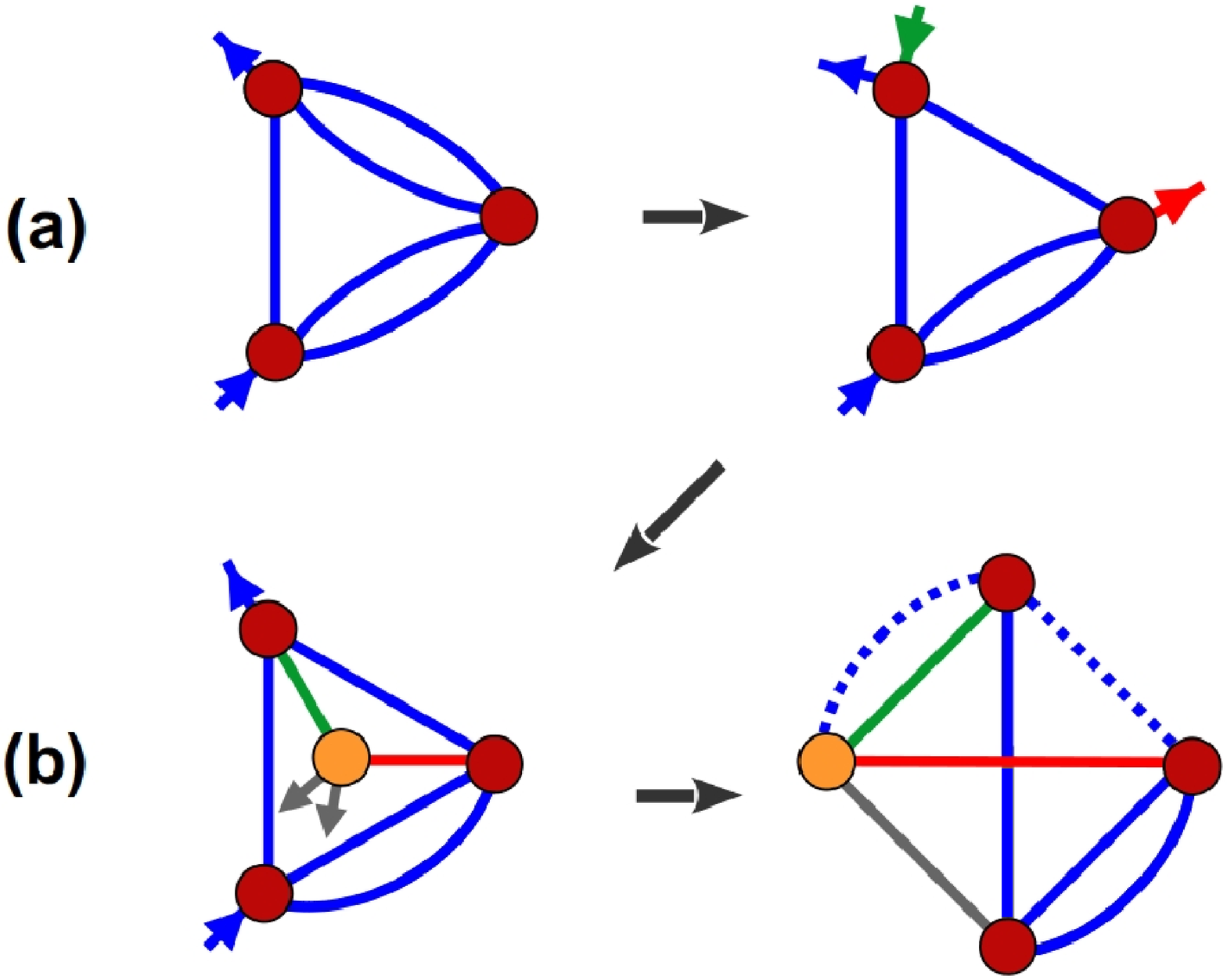} }
{{\bf FIG. 7.} {\small The iterative parquet algorithm Eq. (\ref{lw47.0}),
starting from the FLEX self-energy, is incompatible with $\Phi$ derivability.
(a) Differentiation of the self-energy term at third order in the interaction
gives a term in the parquet kernel series.
(b) Iteration of the self-energy in the parquet algorithm must close the
structure from (a) by adding an interaction, avoiding overcounting
of reducible terms. This generates a novel irreducible component
in the parquet series. A final closure generates the linked
correlation-energy diagram of Fig. 6(b),
which is not a legitimate $\Phi$-derivable contribution.
Since $\Phi$ derivability must hold at every order, no level of
iteration of the parquet kernel can fulfill it.
}}
\vskip 0.25cm

The failure of the relation Eq. (\ref{lw47.6}) to vanish has the
more serious implication that a parquet model does not correspond to a
system with a well defined ground-state energy. Securing that would
require a $\Phi_{\rm PQ}$ conforming to the criteria of Baym and Kadanoff.
If such a functional can be constructed to satisfy Eq.  (\ref{lw47.3}),
say, it will not have the canonical Luttinger-Ward form of either
Eqs. (\ref{kII02}) or (\ref{lw47.1}). The question of the existence
of a stable ground-state configuration stays undecided for parquet.

So far we have shown how both $\Phi$-derivable and parquet models fail
to produce forms for the correlation energy that are fully consistent
both with respect to conservation and to crossing symmetry. However,
unlike parquet, $\Phi$ derivability preserves internal consistency
in the sense of Luttinger and Ward,
\cite{lw}
in particular $\Phi$ as an extremum with respect to external perturbations.

Our conclusions on the limits of both parquet and $\Phi$-derivable models
coincide fully with those of the diagrammatic analysis of Smith.
\cite{roger}
That analysis applies as well to more elaborate $\Phi$-derivable structures
beyond the one corresponding to $stu$/FLEX, underwritten by its Kraichnan
Hamiltonian. In his different functional-integral approach,
oriented towards critical behavior, Jani\v{s} 
\cite{janis1,janis2}
likewise remarks on the discrepancy between the parquet kernel's analytical
properties and those obtained from $\Phi$ derivability.

\subsection{Crossing symmetry and response}

Notionally, while crossing symmetry will apply to scattering off an open
system, response analysis concerns a closed system and thus a different
interplay of two-body vertex and one-body self-energy correlations. Any
complement to the extra term $\Gamma''$,
if found neither in $\Gamma$ itself nor in the self-energy insertions
subsumed in the total two-body response, could make no contribution to that
conserving response within its approximating $\Phi$-derivable framework.
If needed for conservation, the counter-term must show up somewhere.
\cite{gllg}

\centerline{
\includegraphics[height=6truecm]{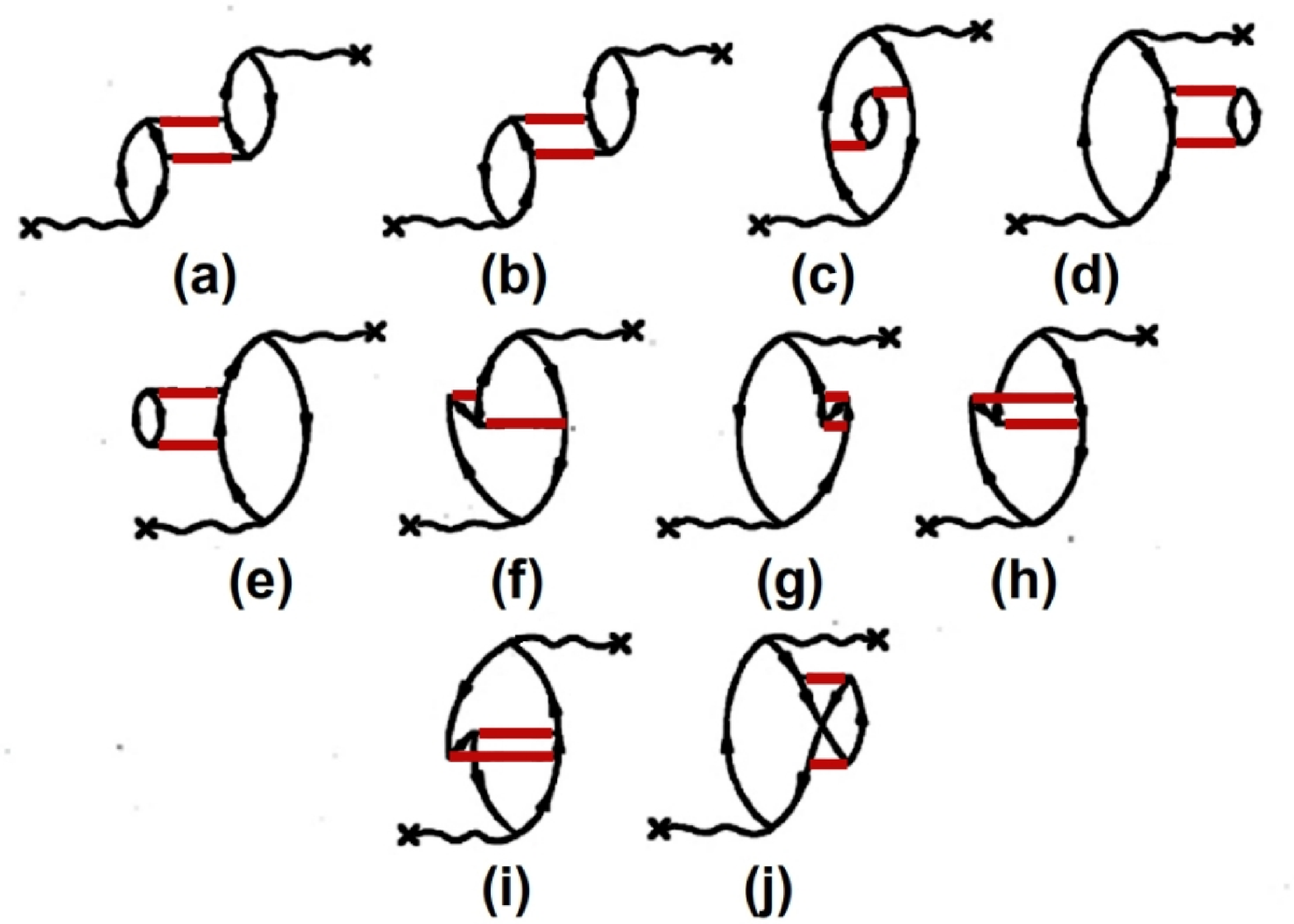}
}
{{\bf FIG. 8.} {\small Damping terms in the conserving
high-frequency summation of the two-body electron-gas polarization
function, exact to second order in $V$ (full horizontal lines),
after Fig. 2 of Glick and Long.
\cite{gllg}
Wavy lines terminating with {\bf x} are couplings to the external
probe, directed lines are free propagators.
Terms (a), (b), (c), (f), (h), and (i) have their kernel
in $\Lambda'$ as generated from $\Phi$. 
For consistency, these two-body vertex components are
summed concurrently with the one-body insertions
(d), (e), (g), and (j) that come from the uncorrelated
bubble $GG$. The overall topology in terms of bare lines
does not discriminate between self-energy and vertex terms,
and its systematic cancellations rely on more than manifest
crossing symmetry.
}}
\vskip 0.25cm

Specializing to the purely computational aspect of the $stu$ and parquet
analyses, we draw attention to Fig. 2 of the paper by Glick and Long,
\cite{gllg}
here replicated in Fig. 8. It
exhibits the dominant high-frequency contributions to the imaginary
(damping) part of the polarization function for the electron gas and
derives from the bare expansion of the density response, generated
by $\delta^2 \Phi/\delta U\delta U$ when the exact correlation energy
is truncated at second order in the bare potential $V$.
Self-energy insertions from the externally coupled propagators
must be computed in systematic superposition with the corresponding
interaction-vertex contributions.

Glick and Long's example demonstrates that, to account systematically
for the dynamical correlations in the response, self-energy contributions
from the propagators external to the two-body interaction $\Gamma$ enter,
as well as those internal to it. This means that a protocol broader than
explicit crossing symmetry determines the bookkeeping that produces the
overall conserving result. In the $\Phi$-derivable approach, a similar
pattern of cancellation also provides the counterbalancing mechanism for
the non-parquet component $\Gamma''$.

The following is of interest. The $\Phi$-derivable model, truncated beyond
second order in $\OV$, reproduces precisely the diagrams of Fig. 8. At second
order, the structure of the kernel $\Lambda$ of $\Phi$ is ambiguously defined
(degenerate, if one likes); it may be envisaged to manifest in any of the
channels $s,t$ or $u$, which is the very reason for forming the composite
K-coupling of Eq. (\ref{kII13}) to avoid overcounting. Nevertheless,
perturbing the system lifts the structural degeneracy, with all three channels
emerging on an equal footing in the Kadanoff-Baym functional derivation of
the second-order kernel $\Lambda'$. In this quite special case $\Gamma'$
is crossing symmetric, yet crossing symmetry is not uniquely assignable to
the generating kernel $\Lambda$. Beyond second order the channel ambiguity
is lifted and crossing symmetry for $\Lambda'$ is lost; but what the
second-order case highlights is that the $stu$ parquet structure is inherent
in $\Phi$ derivability, even if in a weaker sense and
even if insufficient to secure strict crossing symmetry in general.
One can refer to Fig. 5 to see this stated graphically.

Equation (\ref{kII02}) for the correlation energy implies that its
fundamental expansion is in powers of the underlying bare interaction $V$ 
regardless of where it occurs structurally.
This implies in turn that one should look again at the expansion
in terms of the bare propagator $G^{(0)}$ rather than focus
exclusively of the full propagator $G$. As indispensable as $G$ is
as a construct in making sense of the correlation physics,
it tends to hide those instances of $V$ within the propagators that
counterbalance its presence in the skeleton graphs defining the vertex
components; a concealment that, as suggested by Fig. 8, masks how
cancellations pair up among two-body and one-body self-energy elements.

The parquet model's self-energy structures for $G$ are set by crossing
symmetry through the feedback imposed on the self-energy kernel. There,
it is the skeletal topology of $\Gamma$ that governs the processes of
cancellation.
For $\Gamma'$ in the complete $\Phi$-derived two-body Green function,
conservation operates otherwise: as in Fig. 8, competing effects must
cancel in a determinate superposition. It is this that conditions the
topology of the approximate $\Gamma'$, not the other way around.

Since $G$ is an infinitely nested functional of $V$, the
renormalized $\Lambda$ and $\Gamma'$ can well differ diagrammatically
while their bare-expansion analogs, respectively ${\widetilde \Lambda}$
and ${\widetilde \Gamma'}$ will not.
These last two cannot differ in their structure because the
{\em only} topological distinction between the bare graphs of $\Phi$
and the bare graphs of the derived correlated response is the external
perturbation nodes attached to the bare propagators. In other words,
\[
\widetilde \Gamma' = \widetilde \Lambda.
\]
Unlike $\Gamma'$, the linked diagrams of the bare expansion for
$\widetilde \Gamma'$ do not differentiate between one-body and
(two-body) vertex contributions.
They expose the self-energy insertions not only internal to the two-body
Green function but in the external incoming and outgoing lines as well.

By themselves, the internal arrangements of the renormalized four-point
kernel are insufficient for response. The response function's graphs are
closed: it is a contraction of the two-body Green function.
\cite{kb1}
The outer connections of $\Gamma'$ must terminate in two particle-hole
pairs $GG$ to obtain the dynamically correlated contribution.
In the overall accounting the leading uncorrelated particle-hole bubble
$GG$ also plays an explicit role.

The physical response function is the same whether written in
terms of $\widetilde \Gamma'$ or of $\Gamma'$. It follows that
in the latter's renormalized setting the non-parquet component
$\Gamma''$, embedded in the complete response, finds its canceling
counterparts among the self-energies. For standard parquet, despite
the bootstrap Eq. (\ref{lw47.0}), the self-energy terms
in the internal propagating pairs $GG$ are not necessarily tuned to
overall cancellation; crossing symmetry reflects only the skeletal form
of $\Gamma$, not its dynamics. It is an additional assumption that
cancellations in parquet are looked after automatically.
In practice, they are not. Figure 8 gives a clue as to why.

\section{Summary}

We have recovered the parquet equations from an augmented form of
Hamiltonian within Kraichnan's fundamental stochastic embedding
prescription.
\cite{k1,k2}
Our particular re-interpretation of the parquet model inherits
the entire suite of conserving analytic (causal) identities from
the exact many-body description for its generating model Hamiltonian.
Relations that rely explicitly on Fock-space completeness are not
preserved, since Kraichnan averaging must decohere classes of
interaction-entangled multiparticle states
(for example, structures as in Fig. 3).

On the way we have examined the seeming paradox of a fully conserving
pairwise-maximal $\Phi$-derivable theory with crossing symmetric kernel
yet leading to a non-symmetric response kernel on one side
(while still including standard parquet in its structure),
and on the other the pairwise-maximal parquet theory in both elementary
and iterated forms, maintaining crossing symmetry but not conservation.
This prompts thought on which philosophy to follow in
formulating many-body approximations, and for which purpose.

The second lesson of this work goes to a conception of how model correlation
theories operate vis \`a vis the conservation laws in a system closed to
external particle exchange. In understanding fluctuations and response,
the parquet construction can be applied fruitfully within a canonically
founded perspective that respects parquet's pairwise-maximal
topology in logical independence from manifest crossing symmetry,
inherited from the distinct open-system physics of nuclear scattering.

In nuclear scattering, at any rate conceptually,
\cite{zqt}
free fermionic constituents arrive from asymptotic infinity to
encounter an open assembly of the same species. They interact strongly and
the free final products scatter off to infinity. One then expects
the outcome to be governed by the optical theorem,
crossing symmetry and thus the forward-scattering sum rule.
\cite{bb2}

In a setting such as transport, the problem involves constituents
that are always confined to the medium, interacting collectively
and strongly while coupling weakly to an external perturbing probe.
A closed scenario interrogates the system very differently.
Accounting of the self-energy contributions from the initial and
final particle-hole $GG$ pairs as well as the uncorrelated bubble $GG$,
coupled via the probe, now matters, and reflects the main philosophical
difference between standard parquet and its $\Phi$-derivable re-reading
in the ambit of response. The role of the vertex terms demands attention
to systematic counterbalancing from the self-energy terms, including from
incoming and outgoing particle-hole states. Such processes are
assured in $\Phi$-derivability, while in parquet they are assumed.

The elegant application of crossing symmetry to particle-antiparticle
processes, fusing them seamlessly with the less problematic but
structurally disparate particle-particle pair processes,
is a foremost idea in many-body understanding.
For $\Phi$ derivability defined by a Hamiltonian, centered
upon conservation and oriented towards response, one is led to
a violation of crossing symmetry in the derived $stu$ scattering kernel.
In the context of self-energy-versus-vertex accounting, this may
be offset partly through the pattern of mutual
cancellations ensuring conservation.

A fully conserving parquet response theory, no longer crossing symmetric
but sharing the identical pairwise-only arrangement of the original parquet
equations, emerges naturally from the Hamiltonian description of the
$stu$/FLEX approximation. The caveat is that, in it, the
correlation-energy kernel $\Lambda$ and the scattering kernel $\Gamma'$
functionally derived from it remain distinct in playing distinct
roles in the renormalized physics. If their differing structures are
conflated, conservation fails.

The puzzle remains. $\Phi$ derivability in an approximate expansion leads to
crossing-symmetry violations, yet in maintaining conservation it suggests
that the violating components are systematically canceled by other means.
Imposing crossing symmetry on an approximating subset of the two-body
scattering amplitude would seem to take care of systematic cancellation,
yet not in a way that conserves.
Understanding in greater detail just how cancellation acts would therefore
provide a much needed clarification.

How might Kraichnan's idea in itself be taken further?
First, the present analysis is readily extended both to non-uniform
cases and a least to some instances where singular behaviour in any
of the pair channels may break ground-state symmetry.
Applying it to analyze nonperturbative many-body formalisms is
also promising. Variational and coupled-cluster methods
are potential candidates. The stochastic embedding approach
pioneered by Kraichnan may not be the only way to set
approximate many-body approaches on a canonical footing. However,
the power of the method in guaranteeing all the conserving
analytic identities that link one- and two-body correlation functions,
even in approximation and beyond linear response, speak compellingly
for revisiting an original and penetrating analysis long
celebrated in the turbulence-theory community
\cite{frisch}
yet, with rare exceptions,
\cite{kb2}
largely unnoticed by its sister community of many-body theory.

\section*{Acknowledgments}

We acknowledge the support of our respective institutions: for FG, the
University of New South Wales; for TLA, Kent State University.


\begin{thebibliography}{99}
\bibitem{err} 
We point out that Fig. 11 for the induced-interaction approximation in Ref.
\cite{KI}
incorrectly shows its corresponding density response function with $\Gamma$
replacing $\Gamma'$, which is not as required by Eq. (\ref{kII18.5}) of the
present paper.
\end{thebibliography}
\end{document}